\documentclass[twocolumn,showpacs,prb,amsmath,amssymb,aps]{revtex4}

\usepackage{graphicx}
\usepackage{dcolumn}
\usepackage{bm}
\usepackage{epsfig}

\begin{document}


\title{Thermoelectric effects of an Aharonov-Bohm interferometer 
with an embedded quantum dot in the Kondo regime}

\author{Tae-Suk Kim$^{1}$ and S. Hershfield$^{2}$}
\affiliation{$^1$ School of Physics, Seoul National University, 
  Seoul 151-742, Korea \\
 $^2$ Department of Physics, University of Florida, Gainesville FL 32611-8440}

\date{\today}

\begin{abstract}
Thermoelectric effects are studied in an Aharonov-Bohm (AB)
interferometer with an embedded quantum dot in the Kondo regime. 
The AB flux-dependent transmission probability has an asymmetrical shape
arising
from the Fano interference between the direct tunneling path
and the Kondo-resonant tunneling path through a quantum dot. 
The sign and magnitude of thermopower can be modulated by the AB flux
and the direct tunneling amplitude. In addition, the thermopower is 
anomalously enhanced by the Kondo correlation in the quantum dot
near the Kondo temperature ($T_K$).  
The Kondo correlation in the quantum dot also leads to 
crossover behavior in diagonal transport coefficients
as a function of temperature. 
The amplitude of an AB oscillation in electric and thermal conductances
is small at temperatures far above $T_K$, but becomes enhanced
as the system is cooled below $T_K$.
The AB oscillation is strong in the thermopower and the Lorenz number
within the crossover region near the Kondo temperature.

\end{abstract}

\maketitle

\section{Introduction}
 The phase-coherence of an electron's wave function in mesoscopic systems 
enables one to observe a wide variety of 
interesting phenomena in quantum physics. 
One of them is the phase shift experienced by electrons 
due to scattering centers. 
Although the phase cannot be measured directly in bulk systems, 
mesoscopic systems provide an opportunity to 
attain such information. 
Recently the transmission phase shift experienced by electrons 
passing through a quantum dot was measured 
using the Aharonov-Bohm (AB) interferometers
\cite{closering,openring1,openring2,openring3}.
In addition to the direct observation of the phase, 
phase-coherence leads to such phenomena as the 
magnetic and electrostatic AB effects, Fano interference, persistent currents,
universal conductance fluctuations, etc.

 Quantum dots provide a unique opportunity to study nonequilibrium 
many-body effects due to the strong Coulomb interaction and 
quantization of energy levels, spin, and charge.
A Fano resonance was observed\cite{fanodotexpI} in the conductance 
in transport through a quantum dot which is connected to two leads.
The opening of direct tunneling between two leads was believed to be 
the main reason for the observation of the Fano interference pattern. 
The interference between two current paths, resonant and direct tunneling,
leads to the asymmetrical Fano resonance in the differential conductance.
However, the nature of the direct tunneling
path is not clear and the direct tunneling cannot be controlled
in this work\cite{fanodotexpI}. 
Inserting a quantum dot in an AB interferometer, the Fano shape
in the $IV$ curves was also observed\cite{fanodotexpII}. 
Free control of the direct tunneling probability 
is possible using a gate voltage.

 In addition, the asymmetrical shape of a Fano resonance has been observed in 
other 
nanoscopic systems. When magnetic adatoms are deposited on a metallic surface,
the local electronic density of states (DOS) on a metallic surface is modified
due to the Kondo effect occurring at the site of a magnetic adatom. 
Using a scanning tunneling microscope (STM)\cite{adatom1,adatom2}, 
the differential conductance between the STM tip and the metallic surface 
was observed to be of asymmetrical Fano resonance type 
close to the magnetic adatom at low temperature.
Confining electrons to an ellipse\cite{qcorral} in a 
quantum corral, the phase coherence
of an electron's wave functions was detected by imaging the metallic surface 
with STM spectroscopy. 
When a magnetic adatom was
positioned at one of the two foci of the ellipse, 
the STM experiments showed the 
coherently reproduced image of the magnetic atom at the other empty 
focus point.  A Fano interference pattern was also observed in the carbon 
nanotubes\cite{nanotube} 
when Co ions were deposited on the surface of carbon nanotubes. 
The ability to observe the Fano resonance shape in all these experiments depends
on the phase coherence of an electron's wave functions
propagating along a metallic surface.

\begin{figure}
\resizebox{0.45\textwidth}{!}{\includegraphics{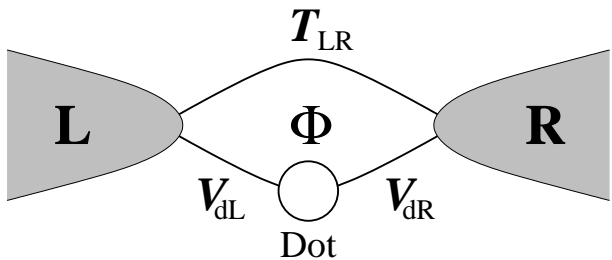}}
\vskip 0.5cm
\caption{Schematic display of Aharonov-Bohm(AB) interferometer.
Electrons can flow from one lead to the other through two paths, 
direct tunneling ($T_{LR}$) and a resonant tunneling via a quantum dot
($V_{dL}, V_{dR}$). The magnetic flux $\Phi$ is threading through 
the AB ring formed of two current paths. The AB phase $\phi = 2\pi \Phi/\Phi_0$
is embedded in the tunneling matrices by the overall phase of  
$V_{dL} T_{LR} V_{Rd} = |V_{dL} T_{LR} V_{Rd}|~e^{i\phi}$.
\label{fanodot}}
\end{figure}

 In this paper, we consider a phase-coherent Aharonov-Bohm interferometer
with an embedded quantum dot (see Fig.~\ref{fanodot}) 
and study theoretically the thermoelectric effects of this system 
\cite{noint}
when the embedded quantum dot lies in the Kondo regime.
The quantum dot in the Kondo regime\cite{expkondo1,expkondo2,expkondo3,thekondo}
 will be called a Kondo dot. 
Though the issue of the transmission phase in this system 
is a very interesting subject, we 
focus on the effects of phase-coherence on the transport coefficients.
The phase-coherence enables us to study the effects of 
the Aharonov-Bohm oscillations as well as the Fano 
interference on the thermoelectric properties of the model system. 
We use the nonequilibrium Keldysh Green's function method and 
the noncrossing approximation for our model study.

 There are two possible paths for the flow of electrons in 
our AB interferometer. The direct tunneling amplitude between two leads is 
a temperature-independent constant, while the tunneling through 
the Kondo dot is strongly temperature dependent due to the Kondo effect.
Different physics is expected above and below the Kondo temperature ($T_K$),
due to the crossover behavior in the Kondo dot as a function of
temperature.
Since the flow of electrons through a quantum dot 
is blocked by the strong repulsive Coulomb interaction
(Coulomb blockade) at high temperatures above $T_K$, 
practically no Fano interference arises and 
the AB oscillations are very {\it weak} in transport coefficients.  
With lowering temperature below $T_K$, the Kondo resonance peak in a quantum dot 
develops close to the Fermi level. 
This opening of an additional resonant current path leads to 
the Fano interference with the direct tunneling path.
Fano interference and AB flux modify the shape of the transmission
probability through the system near the Fermi level.

Out of all the transport coefficients, thermopower is one of the most 
sensitive
to the shape or the particle-hole asymmetry of the transmission probability
and is an appropriate experimental probe to investigate the Fano interference
and the AB effect.
We find that the sign and magnitude of the thermopower can be modulated
by controlling the {\it Aharonov-Bohm phase} with
magnetic fields and 
the {\it tunneling matrices} with varying gate voltages. 
In addition, we find that 
the Kondo correlation enhances the amplitude of AB oscillations 
in diagonal transport coefficients at low $T$ compared to $T_K$. 
Since two diagonal transport coefficients, 
the electric and thermal conductances, 
are influenced by Fano interference and AB flux in the same way, 
their ratio, the Lorenz number, is insensitive to AB flux near the zero 
temperature and is fixed at the Sommerfeld value.
The amplitude of AB oscillations in the thermopower and Lorenz number is 
strong near the Kondo temperature. 
A short paper which presents some of these results appeared elsewhere
\cite{kimselmanPRL}.

 This paper is organized as follows. 
In Sec. \ref{form_sect}, the model Hamiltonian
is introduced and the formulation of electric and heat currents is summarized
leaving the details to Appendix \ref{gdosband} and \ref{wideband}. 
In Sec. \ref{nca_sect}, the modified noncrossing approximation (NCA)
is briefly introduced for our model system. Numerical results from solving 
the NCA 
equations self-consistently are presented in Sec. \ref{result_sect} 
and a conclusion is included in Sec. \ref{sum_sect}.

\section{\label{form_sect} Model and Formalism}
 In this section, we formulate the electric and heat currents in the 
Aharonov-Bohm interferometer which contains a quantum dot in one arm. 
The model system is shown in Fig.~\ref{fanodot}.
Using the Keldysh's nonequilibrium Green's function method
\cite{neqgreen,langreth}, 
the currents are derived in a Landauer-B\"{u}ttiker form. 
The phase-coherence of the model system enables us to 
study the Kondo effects of a quantum dot, 
the Fano interference between the direct path and the Kondo-resonant path, 
and the Aharonov-Bohm effects by applying magnetic fields.

 When the quantum dot lies in the Kondo regime or when there is an odd number of
electrons within a dot and the spin of a highest-lying electron is unpaired,
the quantum dot can be effectively considered 
as a magnetic impurity with spin $\frac{1}{2}$.  
The highest-lying electron can then be described by the Anderson impurity model 
treating other electrons as a background. 
The model Hamiltonian may be written as $H = H_0 + H_1$ where 
\begin{subequations}
\begin{eqnarray}
H_0 &=& \sum_{p=L,R} \sum_{\vec{k}\alpha} \epsilon_{p\vec{k}} 
    c_{p\vec{k}\alpha}^{\dag} c_{p\vec{k}\alpha}^{\phantom{*}} \nonumber\\
 && + \epsilon_d \sum_{\alpha=\uparrow,\downarrow} 
      d_{\alpha}^{\dag} d_{\alpha}^{\phantom{*}} 
  + U n_{\uparrow} n_{\downarrow}, \\
H_1 &=& \frac{1}{V} \sum_{\vec{k}\vec{k}'\alpha} \left[ T_{LR} ~
     c_{L\vec{k}\alpha}^{\dag} c_{R\vec{k}'\alpha}^{\phantom{*}} + H.c. \right] 
    \nonumber\\
  && + \frac{1}{\sqrt{V}} \sum_{p=L,R} \sum_{\vec{k}\alpha} \left[ 
     V_{pd} c_{p\vec{k}\alpha}^{\dag} d_{\alpha}^{\phantom{*}} + H.c. 
     \right]. 
\end{eqnarray}
\end{subequations}
$H_0$ describes the isolated system of a dot and leads and the tunneling between 
a dot and two leads is described by the tunneling Hamiltonian $H_1$. 
Here $c_{p\vec{k}\alpha}$ is the electron annihilation operator of spin direction
$\alpha$ in the $p=L$ (left) $p=R$ (right) lead. 
$d_{\alpha}$ is the annihilation operator
of the highest-lying electron with unpaired spin in a quantum dot.
$T_{LR}$ is the direct tunneling 
matrix from the right lead to the left one 
and $V_{dp}$ is the hopping amplitude 
from the left or right electrode to a quantum dot($d$). 
In our model study,
the wave vector dependence of these tunneling matrices is neglected.

 Since electrons can flow from left to right in two different 
paths (direct tunneling and resonant tunneling via a quantum dot), 
quantum interference is expected when the phase coherence of electrons
is retained. 
Applying a magnetic field to the system, we can study the Aharonov-Bohm(AB) effect
on the transport. 
The AB phase $\phi = 2\pi \Phi/\Phi_0$ is incorporated into the tunneling
matrices 
in a manner that $V_{dL} T_{LR} V_{RL} = |V_{dL} T_{LR} V_{RL}| e^{i\phi}$. 
$\Phi$ is the magnetic flux passing through the system as shown in 
Fig.~\ref{fanodot} and $\Phi_0 = hc/e$ is the flux quantum. 
Applied magnetic fields are assumed to be not strong enough to split
the doubly degenerate spin states of a quantum dot.

\begin{table*}
\caption{\label{table1} New parameters are tabulated in this table.}
\begin{ruledtabular}
\begin{tabular}{cccc}
 &  $(a)$ 
   &  $Z_L^{r} (\epsilon; \phi)$  
    =  $\frac{ \overline{N}_L (\epsilon) }
      { D^r(\epsilon) |D^r(\epsilon)| } 
   \left[ \Gamma_L + \gamma\Gamma_R [g_R^{r} (\epsilon)]^2  
     + (z + z^*) g_R^{r} (\epsilon) \right]$  &  \\
 &  $(b)$ 
   &  $Z_R^{r} (\epsilon; \phi)$  
    =  $\frac{ \overline{N}_R (\epsilon) } { D^r(\epsilon) |D^r(\epsilon)| } 
     \left[ \Gamma_R + \gamma\Gamma_L [g_L^{r} (\epsilon)]^2  
        + (z + z^*) g_L^{r} (\epsilon) \right]$  &  \\
 &  $(c)$ 
   &  $\overline{\Gamma}_L(\epsilon; \phi)$   
    =  $\frac{ \overline{N}_L (\epsilon) } { |D^r(\epsilon)|^2 } 
    \left[ \Gamma_L + \gamma\Gamma_R |g_R^{r} (\epsilon)|^2  
    + z g_R^{r} (\epsilon) + z^* g_R^{a} (\epsilon)  \right]$  &  \\
 &  $(d)$ 
   &  $\overline{\Gamma}_R(\epsilon; \phi)$  
    =  $\frac{ \overline{N}_R (\epsilon) }{ |D^r(\epsilon)|^2 } 
    \left[ \Gamma_R + \gamma\Gamma_L |g_L^{r} (\epsilon)|^2  
    + z^* g_L^{r} (\epsilon) + z g_L^{a} (\epsilon)  \right]$  &  \\
\end{tabular}
\end{ruledtabular}
\end{table*}

 The electric current operator can be defined as a change in the number of electrons
per unit time in the left electrode. 
\begin{eqnarray}
\hat{I}_L 
 &=& -e (-\dot{N}_L) 
 ~=~ \frac{e}{i\hbar} ~[N_L, H] \nonumber\\
 &=& \frac{e}{i\hbar} \frac{1}{V} \sum_{\vec{k}\vec{k}'\alpha} 
   \left[ T_{LR} c_{L\vec{k}\alpha}^{\dag} c_{R\vec{k}'\alpha} 
     - T_{RL} c_{R\vec{k}'\alpha}^{\dag} c_{L\vec{k}\alpha} \right]  \nonumber\\
 && + \frac{e}{i\hbar} \frac{1}{\sqrt{V}} \sum_{\vec{k}\alpha} 
   \left[ V_{Ld} c_{L\vec{k}\alpha}^{\dag} d_{\alpha} 
      - V_{dL} d_{\alpha}^{\dag} c_{L\vec{k}\alpha} \right].
\end{eqnarray}
Here $N_L = \sum_{\vec{k}\alpha} c_{L\vec{k}\alpha}^{\dag} c_{L\vec{k}\alpha}$
is the number operator in the left lead. 
The heat current operator is also defined as a change in the thermal energy 
per unit time in the left lead. 
\begin{eqnarray}
\hat{Q}_L 
 &=& - \frac{1}{i\hbar} [ H_L^{'}, H].
\end{eqnarray}
$H_L^{'}$ is the Hamiltonian of the left lead without the chemical potential 
shift due to the source-drain voltage.  
Using the current conservation in a steady state, both currents
can be written in a Landauer-B\"{u}ttiker form in terms of 
the Green's function of a dot.
\begin{eqnarray}
\begin{pmatrix} I_L \cr Q_L \end{pmatrix}
 &=& \frac{2}{ h} \int d\epsilon \begin{pmatrix} -e \cr \epsilon - \mu_L \end{pmatrix} 
   T(\epsilon) \left[ f_L(\epsilon) - f_R(\epsilon) \right]. 
\end{eqnarray}
Here $f_p(\epsilon) = f(\epsilon-\mu_p, T_p)$ is the Fermi-Dirac thermal 
distribution function in the lead $p=L,R$ when each lead is in thermal equilibrium 
at temperature $T_p$. 
$\mu_p = -e V_p$ is the chemical potential shift due to the applied 
source-drain bias voltage. 
$T(\epsilon)$ is the transmission probability spectral function through 
an AB ring and is related to the Green's function $G_d^{r}$ 
of a dot by the equation,
\begin{eqnarray}
\label{tprobgdos}
T(\epsilon)
 &=& T_0 (\epsilon) - \mbox{Im} \left[ Z_{LR} G_d^{r} \right].
\end{eqnarray}
The first term $T_0(\epsilon)$ 
comes from the direct tunneling and is given by the expression.
\begin{eqnarray}
T_0 (\epsilon) 
 &=& \frac{ 4 \gamma \overline{N}_L(\epsilon) \overline{N}_R(\epsilon) } 
     { |D^r(\epsilon)|^2 }.
\end{eqnarray}
Here $\gamma = \pi^2 N_L N_R |T_{LR}|^2$ is the dimensionless measure of direct 
tunneling of electrons between two leads. $N_{L,R}$ is the density of states (DOS)
at the Fermi level of the left, right leads, respectively. The overlined
DOS is normalized such that its value is unity at the Fermi energy.   
$D^r(\epsilon) = 1 - \gamma g_L^{r}(\epsilon) g_R^{r}(\epsilon)$
and the definitions of reduced Green's functions $g_p^{r,a} (\epsilon)$ 
of each lead ($p=L,R$) can be found in Appendix \ref{gdosband}. 
The resonant tunneling through the quantum dot and the interference effect
are all included in the second term of $T(\epsilon)$.
\begin{eqnarray}
Z_{LR}
 &=& \frac{ 4 Z_L^{r} Z_R^{r} } 
    { \overline{\Gamma}_L(\epsilon) + \overline{\Gamma}_R (\epsilon) }.
\end{eqnarray}
In the absence of the direct tunneling, $Z_{LR}$
is reduced to the familiar expression, $4\Gamma_L \Gamma_R /(\Gamma_L+\Gamma_R)$.
The parameters of $Z_{L,R}^{r}$ and $\overline{\Gamma}_{L,R} (\epsilon)$ are
tabulated in the table~\ref{table1}.  
Detailed derivations and definitions of new notations can be found in Appendix
\ref{gdosband}. 
The line-broadening parameter $\Gamma_p = \pi N_p |V_{dp}|^2$ ($p=L,R$) 
measures the hopping rate of 
electrons between the quantum dot and the leads.
The complex number $z = \pi^2 N_L N_R V_{dL} T_{LR} V_{Rd} 
= \sqrt{\gamma \Gamma_L \Gamma_R} ~ e^{i\phi}$ contains the effect of applied 
magnetic fields through the Aharonov-Bohm phase $\phi = 2\pi \Phi/\Phi_0$. 
This combination of tunneling matrices $V_{dL} T_{LR} V_{Rd}$ has the meaning that 
electrons hop off the quantum dot into the right lead, tunnel from the right lead
to the left one via the direct tunneling, and hop back on the quantum dot.  
One complete circulation of an electron's motion along the Aharonov-Bohm ring
picks up the AB phase $\phi$ generated by the enclosed magnetic flux, $\Phi$. 
Note that $Z_{p}^{r}(\epsilon; -\phi) = Z_{p}^{r}(\epsilon; \phi)$ and 
$\overline{\Gamma}_p (\epsilon; \phi) \neq \overline{\Gamma}_p (\epsilon; -\phi)$
for $p=L,R$, but $\overline{\Gamma}_L (\epsilon; -\phi)
 + \overline{\Gamma}_R (\epsilon; -\phi) = \overline{\Gamma}_L (\epsilon; \phi)
+ \overline{\Gamma}_R (\epsilon; \phi)$. These relations under the inversion of 
the magnetic flux leads to the (broken) AB phase symmetry in (out of) equilibrium,
respectively.

 In the wide conduction band limit, 
all the relevant Green's functions of two electrodes 
become energy-independent except for the thermal functions. 
The transmission probability spectral function reduces to 
(see Appendix \ref{wideband} for details)
\begin{eqnarray}
T (\epsilon) 
 &=& \frac{4\gamma} {(1+\gamma)^2} 
   + \frac{ 4 \overline{\Gamma}_L \overline{\Gamma}_R } 
       { \overline{\Gamma}_L +\overline{\Gamma}_R } ~\mbox{Im} G_{d}^r \nonumber\\
 &&  - \frac{1-\gamma } {1 + \gamma} ~
    \frac{4} { \overline{\Gamma}_L +\overline{\Gamma}_R }~
     \mbox{Im} G_{d}^r \left\{ \overline{\Gamma}_L Z_R^{r} 
        + \overline{\Gamma}_R Z_L^{r} \right\}.
\end{eqnarray}
The first term is the transmission amplitude due to the direct tunneling. 
The second term is the expected form for transport through a quantum dot 
with the renormalized hopping rates $\overline{\Gamma}_L$ and $\overline{\Gamma}_R$. 
The third term is the interference between two current paths. 
We can further simplify the transmission probability spectral function $T(\epsilon)$ as
\begin{subequations}
\begin{eqnarray}
\label{tprobflat}
T (\epsilon) 
 &=& T_0 + 2 \overline{\Gamma} \sqrt{g T_0 (1-T_0)} ~\cos\phi ~\mbox{Re} G_d^r 
    \nonumber\\
 &&  + \overline{\Gamma} \left[ T_0 - g (1 - T_0\cos^2\phi) \right]~
     \mbox{Im} G_d^r, \\
T_0 &=& \frac{4\gamma} {(1+\gamma)^2}, ~~~
  g ~=~ \frac{4\Gamma_L \Gamma_R }{ ( \Gamma_L + \Gamma_R)^2 }.     
\end{eqnarray}
\end{subequations}
Here $ \overline{\Gamma}$ is equal to $\overline{\Gamma}_L +\overline{\Gamma}_R
 = (\Gamma_L + \Gamma_R)/(1+\gamma)$. 
Note that $T_0$ is the transmission probability through the direct tunneling,
and $g$ is the maximum dimensionless linear conductance through a quantum dot in the
absence of the direct tunneling.
This dimensionless conductance also provides 
measures of the asymmetry in the coupling of 
a quantum dot to the left and right electrodes.

 In the linear response regime, we can expand the electric and heat currents 
up to the linear terms of $\delta V = V_L - V_R$ and $\delta T = T_L - T_R$.
The transport coefficients $L_{ij}$ are defined by the relations,
\cite{sivanimry}
\begin{eqnarray}
\begin{pmatrix} I_L \cr Q_L \end{pmatrix} 
 &=& \begin{pmatrix} L_{11} & L_{12} \cr L_{21} & L_{22} \end{pmatrix}
    \begin{pmatrix} V_L - V_R \cr T_L - T_R \end{pmatrix},
\end{eqnarray} 
and can be expressed in terms of the transport integral,
\begin{eqnarray}
I_n (T) &=& \frac{2}{h} \int d\epsilon ~ \epsilon^n T(\epsilon) 
   \left[ -\frac{\partial f }{ \partial \epsilon} \right], 
\end{eqnarray}
as
$L_{11} = e^2 I_0$, $L_{21} = L_{12} T = -e I_1$ and $L_{22} = I_2/T$.
The
linear response conductance, $G = \lim_{V\to 0} dI/dV = L_{11}$,
is given by the equation,
\begin{eqnarray}
G &=& e^2 I_0(T),
\end{eqnarray}
which measures particle-hole symmetrical part of the transmission probability 
$T(\epsilon)$ with respect to the Fermi level.
The thermopower of a quantum dot 
in a two-terminal configuration
can be found in an open circuit($I=0$) by measuring the induced voltage
drop across a quantum dot when the temperature difference between two leads
is applied. The thermopower is defined by the relation
\begin{eqnarray}
S &\equiv& - \lim_{T_L \to T_R} \left. \frac{V_L - V_R }{ T_L - T_R} \right|_{I=0},
\end{eqnarray} 
and can be expressed as
\begin{eqnarray}
S &=& \frac{L_{12} }{ L_{11}}
 ~=~ - \frac{k_B }{ e} \frac{I_1 }{ k_B T  I_0}. 
\end{eqnarray}
Here the constant $k_B/e$ is approximately $86.17$ $\mu$V/K. 
In most metals, the thermopower is of the order of a few $\mu$V/K. 
Since the integral $I_1$ measures the first moment of energy $\epsilon$ 
in the transmission probability, the thermopower probes the particle-hole 
asymmetric part in $T(\epsilon)$ with respect to the Fermi energy.
The
thermal conductance $\kappa$ through an AB ring can be expressed in terms of 
these transport integrals as
\begin{eqnarray}
\kappa 
 &=& \frac{1}{T} \left[ I_2 - \frac{I_1^2}{I_0} \right].
\end{eqnarray}
The
thermal conductance probes the particle-hole symmetric part of the transmission
probability.

\section{\label{nca_sect} Non-crossing approximation}
 To study the thermoelectric effects of our model system, 
we have to find the Green's function, $G_d$, of a quantum dot. 
 Due to the strong repulsive Coulomb interaction in a dot, 
the calculation of $G_d$ is nontrivial and needs a many-body
technique. 
For this purpose, we adopt the noncrossing approximation (NCA), 
which is a self-consistent diagrammatic
method (for details, see Refs.~\cite{eqnca1,eqnca2,neqnca1,neqnca2,kimselman}). 
In our model system, the direct tunneling path renormalizes the Anderson
hybridization functions and leads to a flux-dependent effective 
continuum band for the Anderson impurity (quantum dot), and
the NCA integral equations are accordingly modified as shown below.

 The NCA has been very successful in studying the Anderson impurity models 
in the Kondo regime both in normal metals \cite{eqnca1,eqnca2} and 
out of equilibrium\cite{neqnca1,neqnca2,kimselman}.
The zero-temperature analysis of the NCA self-energies reveals that 
the NCA gives rise to the nonanalytic behavior of the Green's functions
near the Fermi energy. 
However this nonanalytic behavior shows up in the finite temperature NCA 
below the {\it pathological temperature} $T_p$.
The value of $T_p$ can be estimated analytically\cite{eqnca1}
and $T_p \ll T_K$, where $T_K$ is the Kondo temperature,
so that the interesting Kondo effects in physical quantities 
can be computed down to temperatures far below $T_K$ but still
larger than $T_p$.

 Due to the direct tunneling between the two electrodes, 
the effective continuum band for the Anderson impurity (quantum dot) 
is modified from two simple leads. In general, the self-energy for an electron
on an Anderson impurity embedded in an effective electron continuum band can be 
decomposed into two terms, the one-body and many-body contributions,
\begin{subequations}
\begin{eqnarray}
\Sigma_d (\epsilon) 
 &=& \Sigma_c (\epsilon)  + \Sigma_U (\epsilon), \\ 
\Sigma_c (\epsilon)
 &=& \int \frac{d\zeta}{\pi} 
   \frac{\Gamma_{\rm eff} (\zeta)} {\epsilon - \zeta + i\delta}. 
\end{eqnarray}
\end{subequations}
Here $\Sigma_U(\epsilon)$ is the self-energy due to the on-site Coulomb interaction, 
while the first term $\Sigma_c$ comes from hopping into the continuum band. 
In our model system, 
the one-body contribution $\Sigma_c(\epsilon)$ is given by the Eq.(\ref{oneself})
and can be written as 
\begin{eqnarray}
\Sigma_c(t,t') 
 &=& \frac{1}{V} \sum_{p=L,R}\sum_{\vec{k}\vec{k}'} 
    V_{dp} \overline{G}_p (\vec{k}t, \vec{k}'t') V_{pd} \nonumber\\
 && + \frac{1}{ V} \sum_{p=L,R}\sum_{\vec{k}\vec{k}'} 
    V_{dp} G_{p\bar{p}0} (\vec{k}t, \vec{k}'t') V_{\bar{p}d}.
\end{eqnarray}
The
two auxiliary Green's functions, $\overline{G}_p$ and $G_{p\bar{p}0}$,
are introduced in Appendix~\ref{gdosband}. 
This one-body self-energy shows up
as the continuum band propagator in the Feynman diagrams of the 
pseudoparticle Green's functions, which will be defined below.

\begin{figure}
\resizebox{0.45\textwidth}{!}{\includegraphics{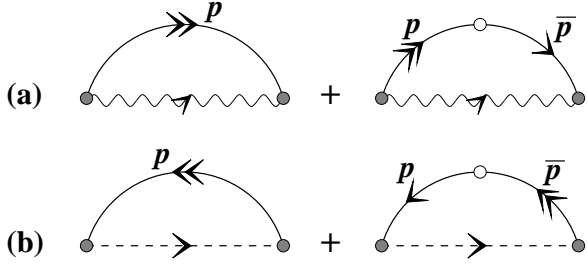}}
\vskip 0.3cm
\caption{NCA self-energy diagrams for (a) a pseudofermion and (b) a slave boson.
The dashed (wavy) lines represent the propagators of pseudofermion (slave boson),
respectively. The solid lines (look at Fig.~\ref{dgreen} for definitions)
are the effective continuum to the Anderson
impurity model and represent all the multiple tunnelings between two leads
($p=L,R$). 
Open (solid) circle denotes the tunneling matrix between two leads 
(between a quantum dot and leads), respectively.
\label{ncaself}}
\end{figure}

 In this paper, we take the limit of $U \to \infty$ and remove 
the double occupancy in the highest-lying electron's state.
Since we are interested in the Aharonov-Bohm effect and the Kondo effect 
in the thermoelectric transport coefficients, the essential physics
can be obtained using the infinite $U$ NCA approach.
In general the vertex correction in the self-energy 
is needed to get the correct Kondo energy scale\cite{vertex} 
when all three Fock spaces of an electron in a dot (empty, singly occupied, and 
doubly occupied configurations) are included.

 In the infinite $U$ NCA, pseudoparticle Green's functions are introduced for 
the empty and singly occupied states,
 and the self-energy equations for these propagators
are solved self-consistently to second order in $V_{dL}$ and $V_{dR}$. 
The empty state is represented by a slave boson operator $b$ 
and singly occupied states (doubly degenerate) are denoted 
by the pseudofermion operator $f_{\alpha}$.
The highest lying electron's annihilation operator in a dot
can then be represented as 
\begin{eqnarray}
d_{\alpha} &=& b^{\dag} f_{\alpha}.
\end{eqnarray} 
The NCA self-energy diagrams for a pseudofermion and a slave boson are displayed 
in Fig.~\ref{ncaself} for our model system and their equations are
\begin{subequations}
\begin{eqnarray}
\Sigma_f (t,t') &=& i\hbar G_b (t,t') \Sigma_c(t,t'), \\
\Sigma_b (t,t') &=& - i\hbar ~N_s ~G_f (t,t') \Sigma_c(t',t), 
\end{eqnarray}
\end{subequations}
respectively.
Here $N_s=2$ accounts for the two spin directions or the degeneracy of pseudofermion
state. $G_{b}$ and $G_{f}$ are the Green's functions of the slave boson and 
pseudofermion operators and are defined as
\begin{eqnarray}
i\hbar G_b (t,t') 
 &=& \langle T_c b(t) b^{\dag}(t') \rangle, \\
i\hbar G_f (t,t') 
 &=& \langle T_c f_{\alpha}(t) f_{\alpha}^{\dag} (t') \rangle. 
\end{eqnarray} 
Projecting
onto the physical Hilbert space\cite{kimselman}, 
the lesser and retarded self-energies can be shown to be given by
the equations,
\begin{subequations}
\begin{eqnarray}
\label{ncaeqns}
\Sigma_f^{r} (\epsilon) 
 &=& \int \frac{d\zeta }{ 2\pi} G_b^{r} (\epsilon - \zeta) \Sigma_c^{>} (\zeta), \\
\Sigma_f^{<} (\epsilon) 
 &=& - \int \frac{d\zeta }{ 2\pi} G_b^{<} (\epsilon - \zeta) \Sigma_c^{<} (\zeta), \\
\Sigma_b^{r} (\epsilon) 
 &=& N_s \int \frac{d\zeta }{ 2\pi} G_f^{r} (\epsilon + \zeta) \Sigma_c^{<} (\zeta), \\
\label{ncaeqnf}
\Sigma_b^{<} (\epsilon) 
 &=& - N_s \int \frac{d\zeta }{ 2\pi} G_f^{<} (\epsilon + \zeta) \Sigma_c^{>} (\zeta).
\end{eqnarray}
\end{subequations}
These four equations are our NCA integral equations to be solved. 
 The lesser and greater self-energies of a quantum dot due to the hopping 
into the two leads are 
\begin{subequations}
\begin{eqnarray}
\Sigma_c^{<} (\epsilon) 
 &=& 2 \overline{\Gamma}_L (\epsilon; -\phi) f_L (\epsilon) 
     + 2 \overline{\Gamma}_R (\epsilon; -\phi) f_R (\epsilon), \\
\Sigma_c^{>} (\epsilon) 
 &=& 2 \overline{\Gamma}_L (\epsilon; -\phi) \bar{f}_L (\epsilon) 
     + 2 \overline{\Gamma}_R (\epsilon; -\phi) \bar{f}_R (\epsilon), 
\end{eqnarray} 
\end{subequations}
where the Anderson hybridization functions are given by the equations,
\begin{subequations}
\begin{eqnarray}
\overline{\Gamma}_L (\epsilon; -\phi)
 &=& \frac{\overline{N}_L (\epsilon) }{ |D^r(\epsilon)|^2 } 
  \left[ \Gamma_L + \gamma\Gamma_R |g_R^{r} (\epsilon)|^2  \right. \nonumber\\
  && \hspace{1.0cm} \left. 
    + z^* g_R^{r} (\epsilon) + z g_R^{a} (\epsilon)  \right], \\
\overline{\Gamma}_R (\epsilon; -\phi)
 &=& \frac{\overline{N}_R (\epsilon) }{ |D^r(\epsilon)|^2 } 
  \left[ \Gamma_R + \gamma\Gamma_L |g_L^{r} (\epsilon)|^2  \right. \nonumber\\
 && \hspace{1.0cm} \left.
  + z g_L^{r} (\epsilon) + z^* g_L^{a} (\epsilon)  \right]. 
\end{eqnarray} 
\end{subequations}
Two Fermi-Dirac functions $f_{L}$ and $f_{R}$ describe the thermal distribution 
in the left and right electrodes, respectively and $\bar{f}_p = 1 - f_p$. 
The Anderson hybridization functions, $\overline{\Gamma}_L$ and $\overline{\Gamma}_R$,
are renormalized due to the direct tunneling term. 
In a wide conduction band limit, the renormalized Anderson hybridization 
becomes independent of energy variable and simplifies to 
\begin{eqnarray}
\overline{\Gamma}_{L,R} (-\phi) 
 &=& \frac{ \left[ \Gamma_{L,R} + \gamma \Gamma_{R,L} 
    \mp 2 \sqrt{\gamma \Gamma_L \Gamma_R} \sin\phi \right] } {(1+\gamma)^2}.
\end{eqnarray}
This reduction of couplings
between the quantum dot and the two leads results in the smaller Kondo 
temperature
than in the absence of the direct tunneling. In addition, the effective 
Anderson hybridization functions are explicitly dependent
on the Aharonov-Bohm phase $\phi$. 
Note that the AB phase dependence of the Anderson hybridization 
functions  is different in the above self-energies 
and in the equations of table~\ref{table1} (c) and (d)
defined in the expression for the transmission probability. 
When a finite source-drain bias voltage is applied, $f_L$ is not equal
to $f_R$, and 
the lesser and the greater self-energies $\Sigma_c^{>,<}$  
do not remain invariant under the inversion of the magnetic flux,
$\phi \to -\phi$. Out of equilibrium, the AB phase symmetry is broken
in the pseudoparticle self-energies and the Green's function of a quantum
dot so that the transmission probabilities are 
not the same under $\phi \to -\phi$. In equilibrium, $f_L = f_R$ and 
the AB phase symmetry under $\phi \to -\phi$ is recovered.

\section{\label{result_sect} Results}
Although we use the formalism presented in Appendix~
\ref{gdosband} which treats electrodes with a general density of states (DOS),
in actual numerical work a Lorentzian DOS is adopted,
\begin{eqnarray}
g_L^{r} (\epsilon) &=& g_R^{r} (\epsilon) ~=~ \frac{D }{ \epsilon + iD}.
\end{eqnarray}
Here $D$ is the bandwidth of two leads and 
is used as the energy unit ($D=1$). 
The formalism for the flat DOS in the leads
or the wide conduction band limit 
is useful in analyzing our numerical results. 
Throughout our numerical works, we use the following set of model parameters
for a quantum dot.
The energy level of a dot is chosen to be $\epsilon_d = -0.5 D$.
The values of $\overline{\Gamma}_L$ and $\overline{\Gamma}_R$ are adjusted 
to satisfy both the total linewidth, 
$\overline{\Gamma}_L + \overline{\Gamma}_R = 0.14 D$, 
and the chosen value of $g$ (the asymmetry coupling factor).

 Since the effective Anderson hybridization function or 
linewidth of the quantum dot gets smaller due to 
the direct tunneling between two leads, 
the Kondo temperature $T_K$ is also suppressed. 
In a wide conduction band limit, 
the linewidth is $\overline{\Gamma} = \Gamma_L (-\phi) + \Gamma_R(-\phi)
= (\Gamma_L + \Gamma_R)/(1+\gamma)$, 
which is independent of the AB phase $\phi$. 
$T_K$ can be estimated by the equation,
\begin{eqnarray}
T_K &=& D \sqrt{c} ~\exp\left(-\frac{1}{c} \right), 
\end{eqnarray}
where the dimensionless exchange coupling 
$c$ is defined by the relation 
$c \equiv {2\overline{\Gamma} }/{ \pi |\epsilon_d|}$. 
One notable point is that the Kondo temperature is very sensitive to 
the value of the direct tunneling probability $T_0$.

 In the NCA approach to the Anderson impurity, the one-body hopping term 
($V_{dL}$ and $V_{dR}$) is used as the expansion parameter. 
Since only a subset of the self-energy diagrams is included up to the infinite
order, the NCA underestimates the one-body contribution ($\Sigma_c$) to 
the self-energy of an impurity site. 
Accordingly, the Fermi liquid relation\cite{langrethfl}
is not satisfied at $T=0$K and the Kondo resonance peak is exaggerated
especially for the orbitally nondegenerate $S=\frac{1}{2}$ Anderson model. 
This may lead to the violation of the causality relation well below $T_K$, 
or to the negative spectral weight 
in other renormalized Green's functions\cite{kimselman,correction}
(the transmission probability $T(\epsilon)$ near $\epsilon=0$ in our case). 
We can remedy this {\it unphysical} situation by exploiting 
the Fermi liquid relation\cite{langrethfl} for the self-energy of a dot,
\begin{eqnarray}
- \mbox{Im} \Sigma_{d}^{r} (\epsilon=0, T=0)
 &=& \overline{\Gamma}.
\end{eqnarray}
This relation follows from the fact that the dot's self-energy 
due to the Coulomb interaction $U$ satisfies the following relation\cite{yamada},
\begin{eqnarray}
- \mbox{Im} \Sigma_{U}(\epsilon, T) 
 &\propto& \epsilon^2 + [\pi k_B T]^2,  
\end{eqnarray}
when $k_BT, |\epsilon| \ll D, U$. 
However the self-energy $\Sigma_{NCA}$ of a dot computed from the NCA 
always satisfies the following inequality,
\begin{eqnarray}
- \mbox{Im} \Sigma_{NCA} (\epsilon=0, T=0)
 &<& \overline{\Gamma}.
\end{eqnarray}
That is, the imaginary part of the NCA self-energy 
is always less than the Fermi liquid value, $\overline{\Gamma}$. 
In fact, the NCA does a relatively good job in producing 
the energy and temperature dependence of the self-energy
near the Fermi energy, but underestimates the one-body
contribution to the dot's self-energy as noted above. 
Note that the one-body contribution 
is temperature-independent, and its form is known once the shape 
of the Anderson hybridization is given. 
We can make corrections in the one-body contribution to the 
NCA self-energy of a dot such that the above Fermi liquid relation 
is qualitatively satisfied. 
With this scheme\cite{kimselman,correction}, 
we remedy the {\it unphysical} behavior of the NCA results at low
temperatures well below $T_K$ in our work.

\begin{figure}
\vskip 0.5cm
\noindent
\resizebox{0.45\textwidth}{!}{\includegraphics{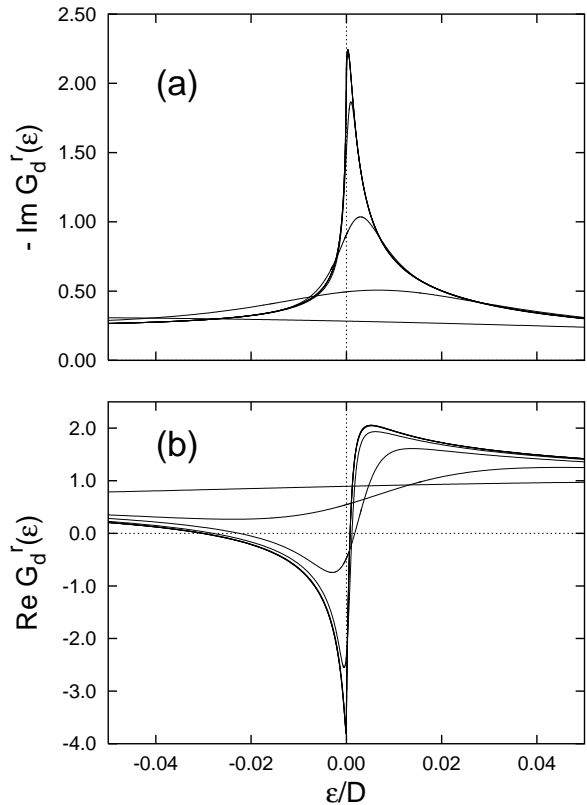}}
\caption{Temperature dependence of a Green's function of a dot near 
the Fermi energy. (a) The imaginary part or the spectral function 
develops the Kondo resonance peak near the Fermi energy with lowering
temperature. (b) The real part develops a dip and peak structure with 
cooling. The temperature $T$ is varied as 
$T/T_K = 100, 20, 4, 0.8, 0.16, 3.2\times 10^{-2}, 6.3\times 10^{-3},
1.3\times 10^{-3}$.  
The last four temperature curves cannot be distinguished with the naked eye.}
\label{kondopeak}
\end{figure}

 {\it Transmission probability spectral function}, $T(\epsilon)$.
 The transmission probability shows the crossover behavior as a function 
of temperature due to the Kondo correlation in the Kondo dot. 
 At high temperatures above $T_K$, electrons flow from the left lead 
to the right one predominantly via the direct tunneling since the tunneling of an electron
from the leads to the dot is prohibited by the strong Coulomb repulsion (Coulomb 
blockade). In this case, the AB effect and the Fano interference 
in the transmission probability are weak. 
As the system is cooled below $T_K$, 
the Kondo resonance peak in the quantum dot develops close to the Fermi level
and provides a new channel for the flow of electrons though the dot. 
The newly opened current path interferes with the direct tunneling path
leading to Fano interference and a strong AB effect
 in the transmission probability.

 The general structure of the transmission spectral functions near 
the Fermi level ($\epsilon=0$) 
can be read off from the equation (\ref{tprobflat}). 
As shown in Fig.~\ref{kondopeak}
$-\mbox{Im} G_d^{r}$ develops the Kondo resonance peak with its width 
of the order of $T_K$ near $\epsilon=0$ while $\mbox{Re} G_d^{r}$ 
varies very rapidly over the energy scale of $T_K$ near $\epsilon=0$ with 
a dip just below $\epsilon = 0$ and a peak above $\epsilon=0$. 
The overall shape of the transmission spectral function $T(\epsilon)$ 
is determined by the value of the AB phase $\phi$ and the sign of $\Delta_c$
[see Eq.~(\ref{tprobflat})],
\begin{eqnarray}
\label{decide}
\Delta_c &\equiv& T_0 - g(1-T_0\cos^2\phi). 
\end{eqnarray}
A typical Fano interference pattern -- a dip and peak
structure -- is expected when $\cos\phi \neq 0$. 
At $\cos\phi = 0$, $T(\epsilon)$ has a dip(peak) resonance structure
if $\Delta_c >(<) 0$, respectively.
Furthermore, we have one exact relation at zero temperature for the transmission 
probability at the Fermi level ($\epsilon =0$), $T(0) = g$, which derives from 
the Fermi liquid relation\cite{langrethfl} for the Green's function of a dot,
$\mbox{Im} G_d^{r} (0) = - 1/\overline{\Gamma}$.

 In equilibrium, $T(\epsilon,\phi) = T(\epsilon, -\phi)$ or
the transmission probability remains invariant 
under the inversion of the magnetic flux, $\Phi \to -\Phi$.
This AB phase symmetry or Onsager relation 
can be deduced from the expression of $T(\epsilon)$ 
in Eq.(\ref{tprobgdos}) or Eq.(\ref{tprobflat}).
If $G_d^{r}(\epsilon)$ remains invariant under the transformation 
$\phi \to -\phi$, so is the transmission probability.
In equilibrium, the Fermi-Dirac thermal distribution functions in two leads 
are identical or $f_L = f_R$.
$G_d^{r}(\epsilon,-\phi) = G_d^{r} (\epsilon,\phi)$ since  
the total Anderson hybridization $\overline{\Gamma}_L(\epsilon;-\phi) +
\overline{\Gamma}_R(\epsilon;-\phi)$ is an even function of $\phi$ and is reduced to 
$\overline{\Gamma} = (\Gamma_L + \Gamma_R)/(1 + \gamma)$ 
in a wide conduction band limit which does not depend on the AB phase. 
Due to the AB phase symmetry and the periodicity in $\phi$, we have only to 
consider $T(\epsilon)$ in the range, $0 \leq \phi \leq \pi$.
Out of equilibrium or when a finite source-drain voltage is applied, 
the two thermal distribution functions are not identical so 
the AB phase symmetry is broken in $G_d$ and $T(\epsilon)$.

 When the direct tunneling is weak, so is the Fano interference, and
the Kondo-related resonance peak persists for the entire range of 
the AB phase. As the direct tunneling is increased, the overall 
magnitude of the transmission probability is also increased,
and the Fano interference becomes more pronounced.

  When $g = 1$ or the quantum dot is coupled symmetrically 
to the left and right leads ($\Gamma_L = \Gamma_R$), 
the transmission probability $T(\epsilon)$ for the case of $T_0 = 0.5$, 
is presented in the short paper\cite{kimselmanPRL}. 
We briefly summarize the shape of $T(\epsilon)$ in this case. 
When $\cos\phi = 0$, the transmission probability has a 
Kondo-related resonance peak near the Fermi level.
When $\cos\phi \neq 0$, the typical shape of a Fano antiresonance 
shows up in the transmission probability. When $0 < \cos \phi < 1$, 
a dip (peak) structure develops below (above) the Fermi level ($E_F$), respectively. 
On the other hand, a dip (peak) structure arises above (below)
$E_F$, respectively, when $-1 < \cos\phi < 0$.

\begin{figure}
\vskip 0.5cm
\noindent
\resizebox{0.45\textwidth}{!}{\includegraphics{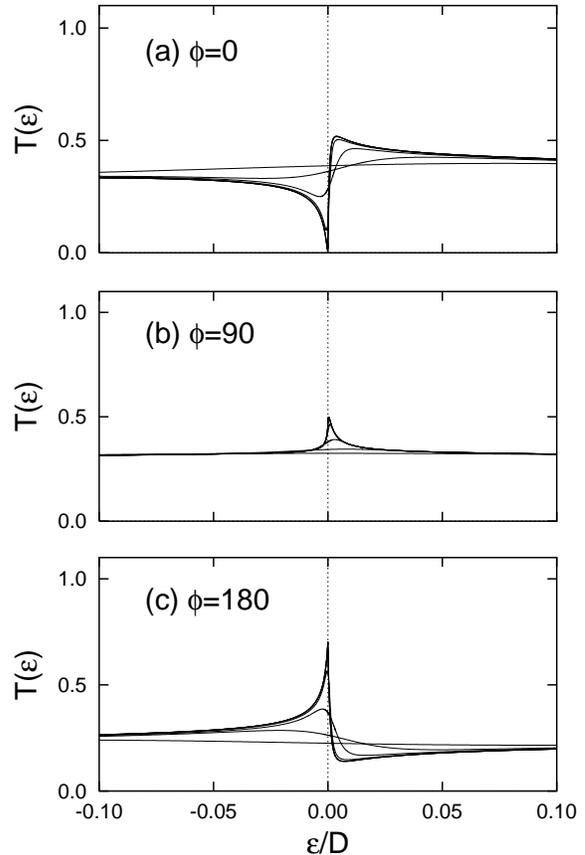}}
\caption{Temperature dependence of the transmission probability
near the Fermi energy for the asymmetrically coupled dot. 
The model parameters are chosen as the asymmetrical factor $g = 0.5$ 
and the direct tunneling probability $T_0 = 0.3$.
 The AB phase is varied as (a) $\phi=0^{\circ}$, (b) $\phi=90^{\circ}$, 
(c) $\phi=180^{\circ}$.}
\label{tprobap}
\end{figure}
\begin{figure}
\vskip 0.5cm
\noindent
\resizebox{0.45\textwidth}{!}{\includegraphics{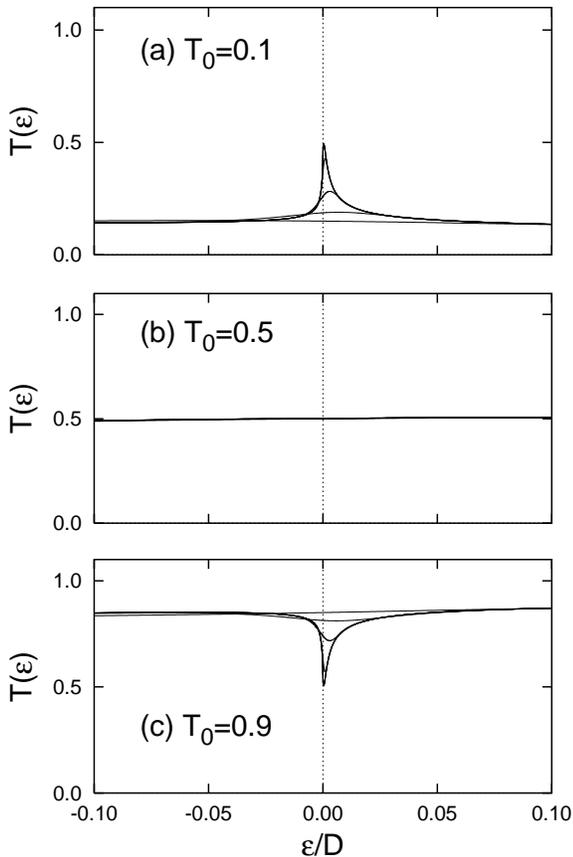}}
\caption{Temperature dependence of the transmission probability
near the Fermi energy for the asymmetrically coupled dot when $\phi=90^{\circ}$. 
The asymmetrical coupling factor is $g = 0.5$ and the direct tunneling 
probability is varied as (a) $T_0=0.1$, (b) $T_0=0.5$, (c) $T_0=0.9$
 from top to bottom panels.}
\label{tprobad}
\end{figure}

 In Figs.~\ref{tprobap} and \ref{tprobad},  we present the computed 
$T(\epsilon)$ when the couplings are asymmetrical or $g < 1$. 
The AB phase dependence of $T(\epsilon)$ is displayed 
in Fig.~\ref{tprobap} when $g = 0.5$ and $T_0 = 0.3$. 
 Comparing the curves in Fig.~\ref{tprobap} and in Fig. 2 in our short
paper\cite{kimselmanPRL}, 
the overall shape of $T(\epsilon)$ is not qualitatively different 
between {\it asymmetrically} 
and {\it symmetrically} coupled dots when $\cos\phi \neq 0$. 
When $\phi=0$ [see Fig.~\ref{tprobap} (a)], $T(\epsilon)$
shows a typical Fano interference pattern, a dip structure below $\epsilon=0$ 
and a peak above $\epsilon=0$. 
With further increasing $\phi$ for $\pi/2 < \phi \leq \pi$, 
the Fano resonance pattern is
inverted with respect to $\epsilon=0$ compared to the case $0 \leq \phi < \pi/2$.
For a symmetrical Kondo resonance peak at $\epsilon=0$ 
or when the energy structure in a quantum dot is particle-hole symmetric, 
exact inversion is expected, $T(\epsilon,\phi) = T(-\epsilon,\pi-\phi)$,
for $0 \leq \phi \leq \pi/2$. 
In our case of the infinite $U$ calculation, the doubly occupied
configuration in the dot is removed from our consideration, and 
particle-hole symmetry is broken. 
Due to this broken particle-hole symmetry, the spectral function of the dot
has more spectral weight in the electron excitations ($\epsilon > 0$)
 than the hole excitations ($\epsilon < 0$)
and the relation $T(\epsilon,\phi) = T(-\epsilon,\pi-\phi)$
for $0 \leq \phi \leq \pi/2$ does not hold in our case,
which can be seen by comparing
the curves in Fig.~\ref{tprobap} (a) and (c).

 When $\cos\phi = 0$ or $\phi=90^{\circ}$, 
$\Delta_c$ is equal to $T_0 - g$ and Re$G_d^{r}$ is absent in 
$T(\epsilon)$. The shape of $T(\epsilon)$ near $\epsilon=0$ is solely determined 
by the sign of $\Delta_c$. 
$\Delta_c$ remains always negative in the case of a {\it symmetrically} coupled dot
so that the Kondo resonance peak persists in the transmission probability 
spectral function over the entire range of the direct tunneling amplitude $T_0$.  
With increasing $T_0$, the Kondo-related peak becomes smaller and 
in the end is buried by the magnitude of $T_0$ as $T_0 \to 1$. 
Note that $T(\epsilon)$ is less than or equal to one for all energies 
in our {\it one transport mode} model. 
The transmission spectral function reaches its maximum possible value $g=1$, 
a unitary Kondo resonance tunneling. 
On the other hand, 
$\Delta_c$ can change its sign from minus to plus in the case of an 
{\it asymmetrically} coupled dot. 
$T(\epsilon)$ retains its peak structure near $\epsilon=0$ 
until $\Delta_c < 0$ [See Fig.~\ref{tprobad} (a)]. 
The maximum value of $T(\epsilon)$ is set by the value of $g$.
When $\Delta_c=0$ or $T_0=g$ [see Fig.~\ref{tprobad} (b)], 
$T(\epsilon)=T_0$ is a constant. 
The interference completely washes out the tunneling
probability through the quantum dot.
The dip structure develops replacing the peak near $\epsilon=0$ as soon as 
$\Delta_c$ becomes positive with increasing direct tunneling amplitude, 
$T_0 > g$ [See Fig.~\ref{tprobad} (c)]. 
The dip structure becomes more pronounced with further increasing $T_0$.

\begin{figure}
\vskip 0.5cm
\noindent
\resizebox{0.45\textwidth}{!}{\includegraphics{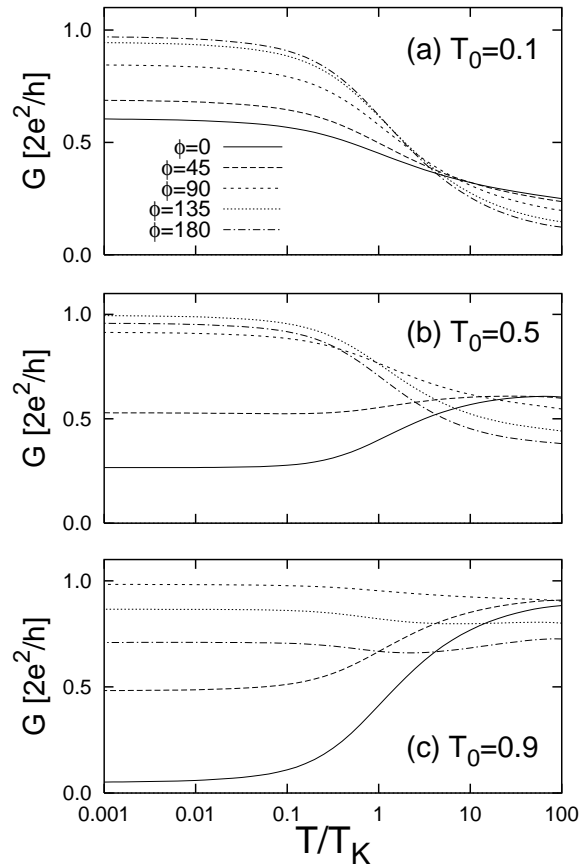}}
\caption{Electric conductance and AB effects. Model parameters: 
$g=1$ (symmetrically coupled dot). Direct tunneling probability 
(a) $T_0 = 0.1$, (b) $T_0=0.5$, (c) $T_0=0.9$ from top to bottom panels.
\label{eqG}}
\end{figure}

 {\it Electric conductance}, $G(T, \phi)$. The temperature and AB phase
dependence of the electric conductance is displayed in Fig.~\ref{eqG}
for a symmetrically coupled dot.
 The temperature dependence of the electric conductance 
is determined by the values of $T_0$ and $\phi$. 
When $T_0$ is small, the electric conductance
is monotonically increasing with lowering temperature for all 
AB phases. When $T_0$ is large, $G(T,\phi)$ is increasing for some range
of $\phi$ and decreasing for others (see Fig.~\ref{eqG}).

 The direct tunneling probability is a temperature-independent constant,
while the transmission through the Kondo dot is strongly temperature
dependent and shows crossover behavior as a function of temperature
due to the Kondo correlation. 
Since the AB effects derive from the interference between two paths, 
the amplitude of the AB oscillation depends on the magnitude of 
the transmission coefficients for the two arms of the AB interferometer.
A stronger direct tunneling amplitude, $T_0$, leads to a
larger amplitude of the AB oscillation.

 The amplitude of the AB oscillation also shows crossover behavior
as a function of temperature. 
The amplitude is weak (strong) at high (low) temperatures above (below) $T_K$, 
respectively. 
This crossover behavior derives from the Fano interference
and the Kondo correlation in the quantum dot. 
Since the quantum dot lies in the Coulomb blockade regime at high temperatures
above $T_K$, 
the Green's function of the dot is negligibly small near the Fermi level
so that the amplitude of the AB oscillation is weak. 
On the other hand, the Kondo correlation with lowering temperature below $T_K$
opens a new transport channel through the quantum dot
and the amplitude of the AB oscillation is enhanced.

\begin{figure}
\vskip 0.5cm
\noindent
\resizebox{0.4\textwidth}{!}{\includegraphics{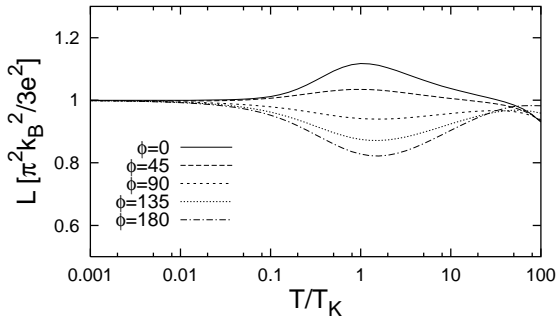}}
\caption{Lorenz number and AB effects. $g = 1$ and $T_0 = 0.5$.
\label{eqL}}
\end{figure}

{\it Thermal conductance and Lorenz ratio}, $\kappa(T,\phi)/T$ and $L(T,\phi)$.
The diagonal transport coefficients, the electric and thermal conductances,
are influenced by the Fano interference and the AB flux in the same way.
The variation of $\kappa(T,\phi)/T$ versus $T$ or $\phi$ is very 
similar to that of electric conductance $G(T,\phi)$.  The Lorenz number $L(T,\phi)$
is defined by the ratio of two diagonal conductances
\begin{eqnarray}
L(T,\phi) &=& \frac{\kappa(T,\phi)}{T G(T,\phi)}.
\end{eqnarray}
The variation of $L(T)$ is displayed in Fig.~\ref{eqL} as a function of 
temperature $T$ and the AB phase $\phi$. 
Our results obey the Wiedeman-Franz law satisfying the relation
$\lim_{T\to 0} L(T,\phi) = \frac{\pi^2}{3} \frac{k_B^2}{e^2}$, 
This relation is independent of the AB flux. 
The AB oscillation in the Lorenz number is strong in the crossover region
or near the Kondo temperature.

\begin{figure}
\vskip 0.5cm
\noindent
\resizebox{0.4\textwidth}{!}{\includegraphics{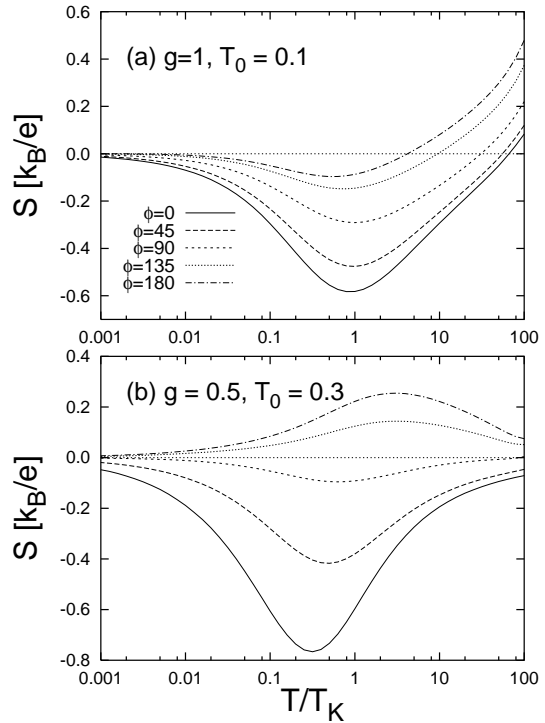}}
\caption{Thermopower and AB effects. (a) $g=1$ and $T_0 = 0.1$,
(b) $g=0.5$ and $T_0 = 0.3$
\label{eqT}}
\end{figure}

{\it Thermopower, $S$}. 
Since thermopower measures the average of the transmission probability
weighted by the electronic excitation energy with respect to the 
Fermi energy, it is a very sensitive probe of the particle-hole
asymmetry in the transmission probability $T(\epsilon)$.
As shown in Figs.~\ref{tprobap} and \ref{tprobad}, 
the shape of $T(\epsilon)$ can be modulated 
through the Fano interference 
by the {\it AB flux} and the {\it direct tunneling amplitude $T_0$}. 
The sign of the thermopower can probe this structure. 
The sign of the thermopower is one of the experimental tools 
which can determine the main carriers of charge and heat. 
When more spectral weight lies in the electron excitations than
in the hole excitations, the main carriers of charge and heat are 
electron excitations, and the sign of $S$ is negative 
[see Figs.~\ref{tprobap}(a) and \ref{eqT}(b)]. 
In the opposite case, the main carriers are hole excitations,
 and the sign of $S$ is positive
[see Fig.~\ref{tprobap}(c) and \ref{eqT}(b)].
In the particle-hole symmetric case (not shown here),
electron and hole excitations carry the same amount of electric and heat 
currents. The signs of electric (heat) current are the same (opposite)
 for electron and hole excitations.
The net result is that the thermopower is zero. 
The particle-hole symmetric case can be realized when the quantum dot
is located in the center of the Coulomb blockade diamond.

 Since the main part of our computed thermopower $S$ in the AB interferometer
is already reported in our short paper\cite{kimselmanPRL}, 
we briefly summarize the main results. 
We find that the sign of $S$ can be changed 
 by the energy structure in a quantum dot\cite{kimselmanPRL,tepI} 
or the gate voltage capacitively coupled to the dot
and by the AB flux\cite{kimselmanPRL}. 
The magnitude of $S$ is anomalously enhanced 
and is of the order of $k_B/e ( \approx 86.17 \mu V/K)$ 
 within the crossover region or near the Kondo temperature. 
For comparison, $S$ is of order $\mu V/K$ in most of normal metals. 
The same enhancement of $S$ is well known in heavy fermion systems 
and in the bulk Kondo systems\cite{eqnca1,eqnca2,kimcox}.

 When the direct tunneling amplitude is weak or $T_0 \ll 1$,
the sign of $S$ remains invariant 
under the inversion of AB flux
at temperatures near or less than $T_K$ [see Fig.~\ref{eqT}(a)]. 
This result follows from the fact that the Fano interference 
is weak and the Kondo-related peak persists in $T(\epsilon)$ 
over the entire range of AB phase $\phi$.
The sign change in $S$ with AB flux can be realized when 
the direct tunneling $T_0$ is increased [see Fig.~\ref{eqT}(b) and 
also Fig. 3 in our short paper\cite{kimselmanPRL}].

 The amplitude of an AB oscillation in the thermopower is strongest within
the crossover temperature region which is similar to the Lorenz number.
The amplitude is weak (strong) 
when the direct tunneling is small (large), respectively
[see Figs.~\ref{eqT}(a) and (b)].

 The sign of $S$ can be changed by varying the direct tunneling amplitude
$T_0$. When $T_0=0.1$ in Fig.~\ref{tprobad}(a), the transmission probability
has more spectral weight in the electron excitations than in the hole
excitations. In this case, the thermopower is negative. 
In Fig.~\ref{tprobad}(b) or when $T_0 = g = 0.5$, $T(\epsilon)$ is
more or less particle-hole symmetric so that $S \approx 0$.
When the direct tunneling is increased or $T_0 = 0.9$ [Fig.~\ref{tprobad}(c)],
holes are the main carriers since the transmission probability is featured
with more spectral weight below the Fermi energy. In this case, the thermopower
is positive.

\section{\label{sum_sect} Summary and Conclusion}
 In this work we studied the effects of Fano interference and 
Aharonov-Bohm phase on the diagonal and off-diagonal transport coefficients
in an AB ring with an embedded quantum dot in the Kondo regime.
The transport properties shows crossover behavior from a
high-temperature regime with no Kondo correlation to a low-temperature
regime with Kondo correlation. 
The crossover behavior manifests itself in the amplitude of 
an AB oscillations 
in the diagonal transport coefficients, the electric and thermal conductances. 
The amplitude is small at high temperatures above $T_K$ but
becomes enhanced below $T_K$ by the Kondo correlation in the quantum dot.
In the case of thermopower, the amplitude of an AB oscillation
is strongest within the crossover region near the Kondo temperature.
The Lorenz number, defined as the ratio of electric and thermal conductances,
is fixed at the Sommerfeld value near zero temperature, but
also shows AB oscillations near the Kondo temperature.

 The effect of the AB flux is more dramatic in the thermopower than 
in diagonal transport coefficients. In addition to the AB 
oscillations in magnitude, even the sign of thermopower can be modulated by 
the AB flux. 
In the case of weak direct tunneling amplitude, the sign change
is not possible under the AB flux since the Fano inference is weak.
When the direct tunneling amplitude is increased,
the AB flux can change the sign of the thermopower.

 Most of studies in quantum dots have focused on the $IV$
curves. Charge confinement is
relatively easy with the use of the gate voltages, but the heat
confinement is a more difficult job possibly because of the easy transfer
of heat via other excitations. 
With more advances in nanotechnology, thermoelectric transport
coefficients will provide additional information about the quantum 
transport in quantum dot systems.

\acknowledgments
 This work is supported in part by the BK21, 
 in part by the National 
Science Foundation under Grant No. DMR 9357474,
and in part by grant No. 1999-2-114-005-5 from the KOSEF.

\appendix

\section{Derivation of equation(\ref{tprobgdos})}\label{gdosband}
 In this section, we derive the expression of electric and thermal currents 
for general shape of the density of states(DOS) for two leads. 
Since the derivation of thermal current is the same as electric current,
we derive the expression of electric current below in detail
and discuss briefly thermal current at the end.

\begin{figure}
\resizebox{0.45\textwidth}{!}
{\includegraphics{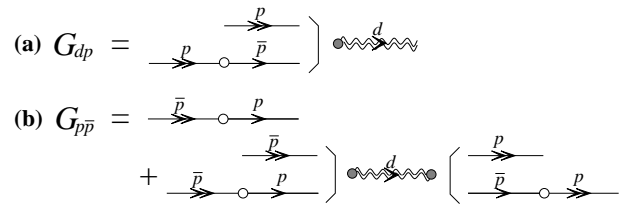}}
\vskip 0.5cm
\caption{Feynman diagrams for the mixed Green's functions which appear in the 
expression of current. Double wavy lines represent the Green's function 
of a quantum dot. The solid line with one arrow means the conduction electron 
propagator in the lead $p=L,R$. 
The solid line with two arrows is the auxiliary Green's function defined in 
 Fig.~\ref{dgreen}(a).  
\label{mgreen}}
\end{figure}
\begin{figure}
\resizebox{0.45\textwidth}{!}
{\includegraphics{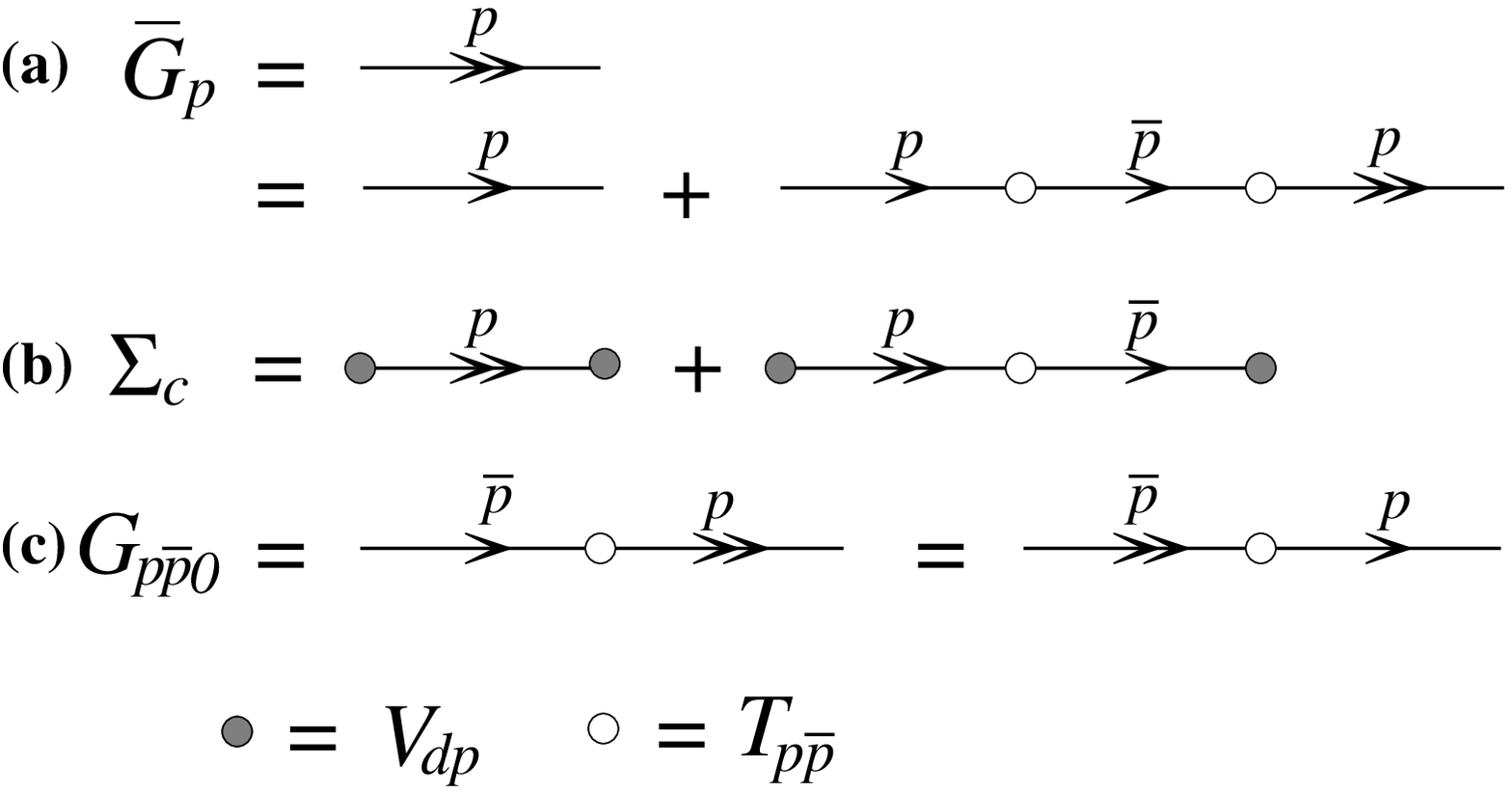}}
\vskip 0.3cm
\caption{Auxiliary Green's functions and one-body self-energy of a dot.
Explaining the bare interaction vertex, the open circle means the direct tunneling 
matrix and the solid circle denotes the hopping amplitude between a quantum dot 
and two leads. 
\label{dgreen}}
\end{figure}

 Using the nonequilibrium Green's function method\cite{neqgreen},
the electric current can be expressed in terms of the lesser mixed Green's functions, 
\begin{eqnarray}
I_L &=& \frac{2e} {h} \frac{1}{V} \sum_{\vec{k}\vec{k}'\alpha} \int d\epsilon ~ 
   \mbox{Im} \left[ T_{LR} ~ G_{RL}^{<} (\vec{k}',\vec{k}; \epsilon) \right] 
   \nonumber\\
  && + \frac{2e}{h} \frac{1}{\sqrt{V}} \sum_{\vec{k}\alpha} \int d\epsilon ~ 
   \mbox{Im} \left[ V_{Ld} ~ G_{dL}^{<} (\vec{k}; \epsilon) \right], 
\end{eqnarray}
where two mixed Green's functions are defined by the equations
\begin{subequations}
\begin{eqnarray}
i\hbar G_{dp} (\vec{k}; t, t') 
 &=& \langle  T d_{\alpha}^{\phantom{*}} (t) c_{p\vec{k}\alpha}^{\dag} (t') \rangle , \\
i\hbar G_{pp'} (\vec{k}t, \vec{k}'t') 
 &=& \langle  T c_{p\vec{k}\alpha}^{\phantom{*}} (t) 
        c_{p'\vec{k}'\alpha}^{\dag} (t') \rangle .
\end{eqnarray}
\end{subequations}
The Feynman diagrams of these two mixed Green's functions are shown 
in Fig.~\ref{mgreen}.
The mixed Green's function $G_{dp}$, shown in Fig.~\ref{mgreen}(a),
is given by the equation
\begin{eqnarray}
G_{dp} (\vec{k}; t, t') 
 &=& \frac{1}{\sqrt{V}} \sum_{\vec{k}_1}
   \int_C dt_1 ~ G_d (t, t_1) V_{dp} \overline{G}_{p} (\vec{k}_1t_1, \vec{k}t').
\end{eqnarray}
The subscript $C$ in the integral symbol denotes the Keldysh contour. 
The auxiliary Green's function $\overline{G}_{p}$ is defined in Fig.~\ref{dgreen}(a) 
and is determined by the following Dyson-like equation
\begin{widetext}
\begin{eqnarray}
\overline{G}_{p} (\vec{k}t; \vec{k}'t') 
 &=& G_{p} (\vec{k}; t, t') ~ \delta_{\vec{k}, \vec{k}'} 
    + \frac{1} {V^2} \sum_{\vec{k}_1 \vec{k}_2} 
    \int_C dt_1 \int_C dt_2  \overline{G}_{p} (\vec{k}t; \vec{k}_1t_1)
    T_{p\bar{p}} G_{\bar{p}} (\vec{k}_2; t_1, t_2) 
    T_{\bar{p}p} G_{p} (\vec{k}'; t_2, t'). 
\end{eqnarray}
\end{widetext}
$G_{p}$ is the Green's function of the conduction electrons in the lead $p$
when $T_{LR} = 0 = V_{dp}$.
\begin{eqnarray}
i\hbar G_{p} (\vec{k}; t, t') 
 &=& \langle  T c_{p\vec{k}\alpha}^{\phantom{*}} (t) c_{p\vec{k}\alpha}^{\dag} (t') 
  \rangle_{0} .
\end{eqnarray}
$G_d$ is the Green's function of a quantum dot, 
\begin{eqnarray}
i\hbar G_{d} (\vec{k}; t, t') 
 &=& \langle  T d_{\alpha}^{\phantom{*}} (t) d_{\alpha}^{\dag} (t') \rangle,
\end{eqnarray}
and its self-energy ($\Sigma_d$) consists of two parts, 
the one-body contribution($\Sigma_c$) coming from tunneling into two reservoirs 
and the many-body contribution ($\Sigma_U$) due to the on-site Coulomb interaction $U$. 
Fig.~\ref{dgreen}(b) gives the one-body contribution to the self-energy 
of a quantum dot. 
\begin{widetext}
\begin{subequations}
\begin{eqnarray}
\Sigma_d(t,t') 
 &=& \Sigma_c(t,t') + \Sigma_U (t,t'), \\
\label{oneself}
\Sigma_c (t,t') 
 &=& \frac{1}{V} \sum_{p\vec{k}\vec{k}'} V_{dp}
     \overline{G}_{p} (\vec{k}t, \vec{k}'t') V_{pd} 
  + \frac{1}{V} \sum_{p\vec{k}\vec{k}'} \frac{1}{V} \sum_{\vec{k}_1} 
    \int_C dt_1~ V_{dp} \overline{G}_{p} (\vec{k}t, \vec{k}_1t_1) 
     T_{p\bar{p}} G_{\bar{p}} (\vec{k}';t_1,t') V_{\bar{p}d}.
\end{eqnarray} 
\end{subequations}
$\Sigma_c$ includes all the multiple tunneling processes between two leads.
Feynman diagrams of the mixed Green's function $G_{p\bar{p}}$ is shown 
in Fig.~\ref{mgreen}(b) and $G_{p\bar{p}}$ is given by the equation
\begin{subequations}
\begin{eqnarray}
G_{p\bar{p}} (\vec{k}t, \vec{k}'t') 
 &=& G_{p\bar{p}0} (\vec{k}t, \vec{k}'t')  
   + \frac{1}{V} \sum_{\vec{k}_1, \vec{k}_1^{'}} \int_C dt_1 \int_C dt_2 ~ 
     \left[ \overline{G}_{p} (\vec{k}t, \vec{k}_1t_1) ~V_{pd} 
         +  G_{p\bar{p}0} (\vec{k}t, \vec{k}_1t_1) ~V_{\bar{p}d}
     \right]   \nonumber\\
 && \hspace{1.0cm}  \times G_d (t_1, t_2) ~
     \left[ V_{d\bar{p}} \overline{G}_{\bar{p}} (\vec{k}_2t_2, \vec{k}'t')
       + V_{dp} G_{p\bar{p}0} (\vec{k}_2t_2, \vec{k}'t')
     \right], \\
G_{p\bar{p}0} (\vec{k}t, \vec{k}'t')
 &=& \frac{1}{V} \sum_{\vec{k}_1} \int_C dt_1 ~
     \overline{G}_{p}(\vec{k}t, \vec{k}_1t_1) 
     T_{p\bar{p}} G_{\bar{p}} (\vec{k}'; t_1,t'). 
\end{eqnarray}
\end{subequations}
\end{widetext}
Another auxiliary Green's function $G_{p\bar{p}0}$, defined in Fig.~\ref{dgreen}(c),
is introduced to facilitate the algebra.  
Introducing the wave-vector-summed Green's functions, 
\begin{subequations}
\begin{eqnarray}
G_p(t,t') 
 &=& \frac{1}{V} \sum_{\vec{k}} G_p (\vec{k}; t, t'), \\
\overline{G}_p(t,t') 
 &=& \frac{1}{V} \sum_{\vec{k}} \overline{G}_p (\vec{k}t, \vec{k}'t'), \\
G_{p\bar{p}} (t,t') 
 &=& \frac{1}{V} \sum_{\vec{k}} G_{p\bar{p}} (\vec{k}t, \vec{k}'t'), \\
G_{dp} (t,t') 
 &=& \frac{1}{\sqrt{V}} \sum_{\vec{k}} G_{dp} (\vec{k}; t, t'), 
\end{eqnarray}
\end{subequations}
the electric current can be written as 
\begin{eqnarray}
\label{Ieqn}
I_L &=& \frac{2e} {h} N_s \int d\epsilon ~ 
  \mbox{Im} \left[ T_{LR} G_{RL}^{<} (\epsilon) \right]  \nonumber\\
 && + \frac{2e} {h} N_s \int d\epsilon ~ 
  \mbox{Im} \left[ V_{Ld} G_{dL}^{<} (\epsilon) \right]. 
\end{eqnarray}
Here $N_s=2$ is two possible spin directions of conduction electrons
in the reservoirs. 
The wave-vector-summed Green's functions are determined by the following equations, 
\begin{widetext}
\begin{subequations}
\begin{eqnarray}
G_{dL} (t,t') 
 &=& \int_C dt_1 ~ D(t,t_1) \left[ V_{dL} \overline{G}_L (t_1, t') 
       + V_{dR} G_{RL0} (t_1, t') \right], \\
\label{GrnRL}
G_{RL} (t, t') 
 &=& G_{RL0} (t, t') + \int_C dt_1 \int_C dt_2 ~ 
     \left\{ \overline{G}_{R} (t, t_1) ~V_{Rd} + G_{RL0} (t, t_1) ~V_{Ld} \right\} 
      \nonumber\\
 && \hspace{2.5cm} \times G_d (t_1, t_2) ~
     \left\{ V_{dL} \overline{G}_{L} (t_2, t') + V_{dR} G_{RL0} (t_2, t') \right\}
  \nonumber\\
 &=& G_{RL0} (t, t') + \int_C dt_1 ~ 
   \left\{ \overline{G}_{R} (t, t_1) ~V_{Rd} + G_{RL0} (t, t_1) ~V_{Ld} \right\}~
   G_{dL} (t_2,t'), \\
\label{indgrn1}
\overline{G}_L (t,t') 
 &=& G_L(t,t') + \int_C dt_1 \int_C dt_2 ~ \overline{G}_L (t,t_1) T_{LR} 
   G_R (t_1,t_2) T_{RL} G_L(t_2,t'), \\
\label{indgrn2}
G_{RL0} (t,t') 
 &=& \int_C dt_1 ~ G_R (t,t_1) T_{RL} \overline{G}_{L} (t_1, t') 
 ~=~ \int_C dt_1 ~ \overline{G}_R (t,t_1) T_{RL} G_{L} (t_1, t').  
\end{eqnarray}
\end{subequations}
\end{widetext}
$G_{LR}$ is obtained from the equation ($\ref{GrnRL}$) by interchanging 
$L\leftrightarrow R$.

The retarded, advanced, and lesser 
Green's functions of the left and right reservoirs (summed over the wave vector) 
can be written as
\begin{widetext}
\begin{subequations}
\begin{eqnarray}
G_p^{r,a} (\epsilon) 
 &=& \frac{1}{V} \sum_{\vec{k}} 
  \frac{1}{\epsilon - \epsilon_{p\vec{k}} \pm i\delta}
 ~=~ \int d\zeta \frac{ N_p(\zeta) }{ \epsilon - \zeta \pm i\delta }
 ~=~ \pi N_p ~ g_p^{r,a} (\epsilon), \\
G_p^{<} (\epsilon) 
 &=& 2\pi N_p(\epsilon) ~ f_p (\epsilon)
 ~=~ 2\pi N_p ~ \overline{N}_p (\epsilon) ~ f_p (\epsilon)
 ~\equiv~ 2\pi N_p ~F_p(\epsilon), \\
\end{eqnarray}
\end{subequations}
Here $N_p (\epsilon) = N_p \cdot \overline{N}_p (\epsilon)$ is the DOS of two leads
and is normalized such that $\overline{N}_p (0)=1$. Reduced Green's function
$g_p^{r,a}$ is introduced to simplify our algebra below and 
$\overline{N}_p (\epsilon) = - \mbox{Im} g_{p}^r(\epsilon)$. 
The auxiliary Green's functions, given by the equations 
(\ref{indgrn1}, \ref{indgrn2}), are 
\begin{subequations}
\begin{eqnarray}
\overline{G}_p^{r,a} (\epsilon) 
 &=& \frac{G_p^{r,a} (\epsilon) } {D^{r,a} (\epsilon)}
 ~=~ \pi N_p ~\frac{g_p^{r,a} (\epsilon) } {D^{r,a} (\epsilon)},  \\
\overline{G}_p^{<} (\epsilon) 
 &=& \frac{1} {|D^r(\epsilon)|^2}  \left[ G_p^{<} (\epsilon) 
    + |T_{LR}|^2 |G_{p}^{r} (\epsilon) |^2 ~ G_{\bar{p}}^{<} (\epsilon) \right], \\
G_{p\bar{p}0}^{r,a} (\epsilon) 
 &=& \frac{T_{p\bar{p}} } { D^{r,a}(\epsilon)} ~ 
  G_p^{r,a} (\epsilon) G_{\bar{p}}^{r,a} (\epsilon)
 ~=~ \frac{T_{p\bar{p}} } {D^{r,a}(\epsilon)} ~ 
  \pi^2 N_L N_R g_L^{r,a} (\epsilon) g_{R}^{r,a} (\epsilon), \\
G_{p\bar{p}0}^{<} (\epsilon) 
 &=& \frac{T_{p\bar{p}} } { |D^{r}(\epsilon)|^2} ~ 
   \left[ G_p^{<} (\epsilon) G_{\bar{p}}^{a} (\epsilon)
       + G_p^{r} (\epsilon) G_{\bar{p}}^{<} (\epsilon) \right]  \nonumber\\
 &=& \frac{T_{p\bar{p}} } {|D^{r}(\epsilon)|^2} ~ 2\pi^2 N_L N_R 
   \left[ g_{\bar{p}}^{a} (\epsilon) F_p (\epsilon) 
         + g_p^{r} (\epsilon) F_{\bar{p}} (\epsilon) \right].
\end{eqnarray}
\end{subequations}
New variables are introduced,
\begin{subequations}
\begin{eqnarray}
D^{r,a} (\epsilon) 
 &=& 1 - |T_{LR}|^2 G_L^{r,a} (\epsilon) G_{R}^{r,a} (\epsilon)
 ~=~ 1 - \gamma ~ g_L^{r,a} (\epsilon) g_{R}^{r,a} (\epsilon), \\
\gamma &=& \pi^2 N_L N_R |T_{LR}|^2, ~~
 \Gamma_p ~=~ \pi N_p |V_{dp}|^2. 
\end{eqnarray}
\end{subequations}
The electric current [Eq.(\ref{Ieqn})] can be written as 
\begin{subequations}
\begin{eqnarray}
I_L &=& \frac{2e}{h} \int d\epsilon ~ \mbox{Im} G_{IL}^{<} (\epsilon), \\
G_{IL}^{<} (\epsilon) 
 &=& 2 T_{LR} G_{RL}^{<} (\epsilon) 
    + 2 V_{Ld} G_{dL}^{<} (\epsilon)  \nonumber\\
 &=& 2 T_{LR} G_{RL0}^{<} (\epsilon) \nonumber\\
 && + 2 G_d^{a} \{ V_{dL} \overline{G}_L^{a} + V_{dR} G_{RL0}^{a} \} T_{LR}
    \{ \overline{G}_R^{<} V_{Rd} + G_{RL0}^{<} V_{Ld} \}  \nonumber\\
 && + 2 G_d^{r} \{ V_{dL} \overline{G}_L^{<} + V_{dR} G_{RL0}^{<} \} 
    \{ T_{LR} \overline{G}_R^{r} V_{Rd} + (1 + T_{LR} G_{RL0}^{r} ) V_{Ld} \} 
   \nonumber\\
 && + 2 G_d^{<} \{ V_{dL} \overline{G}_L^{a} + V_{dR} G_{RL0}^{a} \} 
    \{ T_{LR} \overline{G}_R^{r} V_{Rd} + (1 + T_{LR} G_{RL0}^{r} ) V_{Ld} \}. 
\end{eqnarray}
\end{subequations}
Note the following reduction of equations
\begin{subequations}
\begin{eqnarray}
V_{dL} \overline{G}_L^{a} + V_{dR} G_{RL0}^{a}
 &=& \frac{ G_{L}^{a} } {D^{r} }  \left[ V_{dL} + V_{dR} G_{R}^{a} T_{RL} \right]
 ~=~ \frac{ G_{L}^{a} } { D^{r} } ~ \overline{V}_{dL}^{a}, \\
1 + T_{LR} G_{RL0}^{r}
 &=& \frac{1} {D^r}, \\
V_{dL} \overline{G}_L^{<} + V_{dR} G_{RL0}^{<}
 &=& \frac{1} {|D^r|^2} \left[ \overline{V}_{dL}^{r} G_{L}^{<} 
       + \overline{V}_{dR}^{r} G_{R}^{<} T_{RL} G_{L}^{a} \right], \\
\overline{G}_R^{<} V_{Rd} + G_{RL0}^{<} V_{Ld}
 &=& \frac{1} {|D^r|^2} \left[ G_{R}^{<} \overline{V}_{Rd}^{a} 
       + G_{R}^{r} T_{RL} G_{L}^{<} \overline{V}_{Ld}^{a} \right], 
\end{eqnarray}
\end{subequations}
where the renormalized tunneling matrices are introduced to simplify the 
equations.
\begin{subequations}
\begin{eqnarray}
\overline{V}_{dp}^{r,a}
 &\equiv&  V_{dp} + V_{d\bar{p}} G_{\bar{p}}^{r,a} T_{\bar{p}p},   \\ 
\overline{V}_{pd}^{r,a}
 &\equiv&  V_{pd} + T_{p\bar{p}} G_{\bar{p}}^{r,a} V_{\bar{p}d}. 
\end{eqnarray}
\end{subequations}
The current spectral function in the left lead can be written as
\begin{eqnarray}
G_{IL}^{<} (\epsilon) 
 &=& \frac{4\gamma} {|D^r|^2}  \left\{ g_{L}^{a} (\epsilon) F_{R}(\epsilon) 
    + g_{R}^{r} (\epsilon) F_{L}(\epsilon) \right\} \nonumber\\
 && + \frac{2} {|D^r|^2} \frac{G_d^{a} } {D^a} 
     \left\{ \overline{V}_{dL}^{a} G_{L}^{a} T_{LR} G_{R}^{<} \overline{V}_{Rd}^{a} 
      + |T_{LR}|^2 G_{L}^{a} G_{R}^{r} \cdot 
         \overline{V}_{dL}^{a} G_{L}^{<} \overline{V}_{Ld}^{a}  
     \right\} \nonumber\\
 && + \frac{2} {|D^r|^2} \frac{G_d^{r} } {D^r} 
     \left\{ \overline{V}_{dL}^{r} G_{L}^{<} \overline{V}_{Ld}^{r} 
      + \overline{V}_{dR}^{r} G_{R}^{<} T_{RL} G_{L}^{a} \overline{V}_{Ld}^{r}  
     \right\} \nonumber\\
 && + \frac{2}{|D^r|^2} ~ G_{d}^{<} \cdot G_{L}^{a} | \overline{V}_{dL}^{a} |^2.
\end{eqnarray}
We also find the current spectral function in the right lead, 
\begin{eqnarray}
G_{IR}^{<} (\epsilon) 
 &=& \frac{4\gamma}{|D^r|^2}  \left\{ g_{R}^{a} (\epsilon) F_{L}(\epsilon) 
    + g_{L}^{r} (\epsilon) F_{R}(\epsilon) \right\} \nonumber\\
 && + \frac{2}{|D^r|^2} \frac{G_d^{a} } {D^a} 
     \left\{ \overline{V}_{dR}^{a} G_{R}^{a} T_{RL} G_{L}^{<} \overline{V}_{Ld}^{a} 
      + |T_{LR}|^2 G_{R}^{a} G_{L}^{r} \cdot 
         \overline{V}_{dR}^{a} G_{R}^{<} \overline{V}_{Rd}^{a} 
     \right\} \nonumber\\
 && + \frac{2} {|D^r|^2} \frac{G_d^{r} }{D^r} 
     \left\{ \overline{V}_{dR}^{r} G_{R}^{<} \overline{V}_{Rd}^{r} 
      + \overline{V}_{dL}^{r} G_{L}^{<} T_{LR} G_{R}^{a} \overline{V}_{Rd}^{r}  
     \right\} \nonumber\\
 && + \frac{2} { |D^r|^2} ~ G_{d}^{<} \cdot G_{R}^{a} | \overline{V}_{dR}^{a} |^2.
\end{eqnarray}
Since only the imaginary parts are relevant to the expression of electric current, 
the current spectral function in each lead can be further simplified.
\begin{subequations}
\begin{eqnarray}
\mbox{Im} G_{IL}^{<} (\epsilon) 
 &=& T_0(\epsilon)  \left\{  f_{R}(\epsilon) - f_{L}(\epsilon) \right\} 
  + 2 G_{d}^{<} \cdot \overline{\Gamma}_L (\epsilon)   \nonumber\\
 && + \frac{2} {|D^r|^2} \mbox{Im} \left\{ \frac{G_d^{r} } {D^r} 
     \left\{ \overline{V}_{dL}^{r} G_{L}^{<} \overline{V}_{Ld}^{r} 
       [ 1 - |T_{LR}|^2 G_{L}^{r} G_{R}^{a} ] 
      + \overline{V}_{dR}^{r} G_{R}^{<} T_{RL} [G_{L}^{a} - G_{L}^{r}]
        \overline{V}_{Ld}^{r}  
     \right\}  \right\}, \\
\mbox{Im} G_{IR}^{<} (\epsilon) 
 &=& T_0 (\epsilon) \left\{ f_{L}(\epsilon) - f_{R}(\epsilon) \right\} 
  + 2 G_{d}^{<} \cdot \overline{\Gamma}_R (\epsilon)  \nonumber\\
 && + \frac{2 } {|D^r|^2} \mbox{Im} \left\{ \frac{G_d^{r}}{D^r} 
     \left\{ \overline{V}_{dR}^{r} G_{R}^{<} \overline{V}_{Rd}^{r} 
       [ 1 - |T_{LR}|^2 G_{R}^{r} G_{L}^{a} ] 
      + \overline{V}_{dL}^{r} G_{L}^{<} T_{LR} [G_{R}^{a} - G_{R}^{r}] 
         \overline{V}_{Rd}^{r}  
     \right\}  \right\}, 
\end{eqnarray}
\end{subequations}
where we introduced the transmission probability $T_0$ for the direct tunneling
and the renormalized Anderson hybridization $\overline{\Gamma}_p$, 
\begin{subequations}
\begin{eqnarray}
T_0(\epsilon) 
 &=& \frac{4\gamma }{ |D^r|^2} \overline{N}_{L}(\epsilon) \overline{N}_{R}(\epsilon), \\
\overline{\Gamma}_p (\epsilon)
 &=& \frac{ \pi N_{p}(\epsilon) | \overline{V}_{dp}^{a} |^2 }
    { |D^r(\epsilon)|^2 }.
\end{eqnarray}
\end{subequations}
To simplify the algebra, we write 
\begin{subequations}
\begin{eqnarray}
\mbox{Im} G_{Ip}^{<} (\epsilon) 
 &=& T_0(\epsilon)  \left\{  f_{\bar{p}}(\epsilon) - f_{p}(\epsilon) \right\} 
  + 2 G_{d}^{<} \cdot \overline{\Gamma}_p (\epsilon) 
  + 2 \sum_{q=L,R} A_{pq} G_q^{<} (\epsilon), \\
A_{LL} 
 &=& \frac{1 }{ |D^r|^2} \mbox{Im} \left\{ \frac{G_d^{r} }{ D^r} 
        \overline{V}_{dL}^{r} \overline{V}_{Ld}^{r} 
       [ 1 - |T_{LR}|^2 G_{L}^{r} G_{R}^{a} ]  \right\}, \\
A_{LR} 
 &=& \frac{1 }{ |D^r|^2} \mbox{Im} \left\{ \frac{G_d^{r} }{ D^r}  
    \overline{V}_{dR}^{r} T_{RL} [G_{L}^{a} - G_{L}^{r}]
        \overline{V}_{Ld}^{r}  \right\}, \\
A_{RL} 
 &=& \frac{1 }{ |D^r|^2} \mbox{Im} \left\{ \frac{G_d^{r} }{ D^r}  
      \overline{V}_{dL}^{r} T_{LR} [G_{R}^{a} - G_{R}^{r}] 
         \overline{V}_{Rd}^{r} \right\}, \\
A_{RR} 
 &=& \frac{1 }{ |D^r|^2} \mbox{Im} \left\{ \frac{G_d^{r} }{ D^r}   
      \overline{V}_{dR}^{r} \overline{V}_{Rd}^{r} 
       [ 1 - |T_{LR}|^2 G_{R}^{r} G_{L}^{a} ]  \right\}.
\end{eqnarray}
\end{subequations}
Using the current conservation in a steady state, $I_L + I_R = 0$, 
the expression of $G_{d}^{<}$ can be obtained in terms of the retarded 
and advanced Green's functions.
\begin{eqnarray}
G_{d}^{<} 
 &=& - \frac{ 1 }{ \overline{\Gamma}_L (\epsilon) + \overline{\Gamma}_R (\epsilon) }
  \left[ ( A_{LL} + A_{RL} ) G_{L}^{<} + ( A_{LR} + A_{RR} ) G_{R}^{<} \right].  
\end{eqnarray}
Inserting the lesser Green's function of a quantum dot into $G_{IL}^{<}$, 
we find the current spectral function,
\begin{eqnarray}
\mbox{Im} G_{IL}^{<} (\epsilon)
 &=& T_0(\epsilon) ~\left[ f_R(\epsilon) - f_L(\epsilon) \right]  \nonumber\\
 && + \frac{ 2 }{ \overline{\Gamma} (\epsilon) }
    \left[ ( \overline{\Gamma}_R A_{LL} - \overline{\Gamma}_L A_{RL} ) G_{L}^{<} 
        + ( \overline{\Gamma}_R A_{LR} - \overline{\Gamma}_L A_{RR} ) G_{R}^{<}
   \right]. 
\end{eqnarray}
After some algebra, we can readily derive the following identities,
\begin{subequations}
\begin{eqnarray}
\overline{\Gamma}_R A_{LL} - \overline{\Gamma}_L A_{RL}
 &=& \frac{1 }{ |D^r|^4} \mbox{Im} \frac{G_{d}^{r} }{ D^r} D^a ~
   \pi N_R(\epsilon) ~\overline{V}_{dL}^r \overline{V}_{Ld}^r  
            \overline{V}_{dR}^r \overline{V}_{Rd}^r, \\
\overline{\Gamma}_R A_{LR} - \overline{\Gamma}_L A_{RR}
 &=& - \frac{1 }{ |D^r|^4} \mbox{Im} \frac{G_{d}^{r} }{ D^r} D^a ~ 
   \pi N_L(\epsilon) ~\overline{V}_{dL}^r \overline{V}_{Ld}^r  
            \overline{V}_{dR}^r \overline{V}_{Rd}^r.
\end{eqnarray}
\end{subequations}
Using these results, the current spectral function is reduced to a simple form,
\begin{subequations}
\begin{eqnarray}
\mbox{Im} G_{IL}^{<} (\epsilon) 
 &=& T(\epsilon) ~\left[ f_R (\epsilon) - f_L (\epsilon) \right], \\
T(\epsilon)
 &=& T_0 (\epsilon) - \mbox{Im} \left[ G_d^{r} Z_{LR} \right].
\end{eqnarray}
\end{subequations}
Newly introduced parameters are given by the equations, 
\begin{subequations}
\begin{eqnarray}
Z_{LR} &=& \frac{ 4 Z_L^{r} Z_R^{r} }
       { \overline{\Gamma}_L(\epsilon) + \overline{\Gamma}_R (\epsilon) }, \\
Z_L^{r} (\epsilon) 
 &=& \frac{\pi N_L(\epsilon) \overline{V}_{dL}^{r} \overline{V}_{Ld}^{r} }
     { D^r(\epsilon) |D^r| }
 ~=~ \frac{\overline{N}_L (\epsilon) }{ D^r(\epsilon) |D^r| } 
   \left[ \Gamma_L + \gamma\Gamma_R [g_R^{r} (\epsilon)]^2  
    + (z + z^*) g_R^{r} (\epsilon) \right], \\
Z_R^{r} (\epsilon) 
 &=& \frac{\pi N_R(\epsilon) \overline{V}_{dR}^{r} \overline{V}_{Rd}^{r} }
      { D^r(\epsilon) |D^r| }
 ~=~ \frac{\overline{N}_R (\epsilon) }{ D^r(\epsilon) |D^r| } 
  \left[ \Gamma_R + \gamma\Gamma_L [g_L^{r} (\epsilon)]^2  
    + (z + z^*) g_L^{r} (\epsilon) \right], \\
\overline{\Gamma}_L(\epsilon)
 &=& \frac{\pi N_{L} (\epsilon) |V_{dL}^{a}|^2 }{ |D^r|^2} 
 ~=~ \frac{\overline{N}_L (\epsilon) }{ |D^r|^2} 
   \left[ \Gamma_L + \gamma\Gamma_R |g_R^{r} (\epsilon)|^2  
    + z g_R^{r} (\epsilon) + z^* g_R^{a} (\epsilon)  \right], \\
\overline{\Gamma}_R(\epsilon)
 &=& \frac{\pi N_{R} (\epsilon) |V_{dR}^{a}|^2 }{ |D^r|^2}  
 ~=~ \frac{ \overline{N}_R (\epsilon) }{ |D^r|^2} 
   \left[ \Gamma_R + \gamma\Gamma_L |g_L^{r} (\epsilon)|^2  
    + z^* g_L^{r} (\epsilon) + z g_L^{a} (\epsilon)  \right].
\end{eqnarray}
\end{subequations}
Here $z = \pi^2 N_L N_R V_{dL} T_{LR} V_{Rd}
= \sqrt{\gamma \Gamma_L \Gamma_R} e^{i\phi}$ where $\phi$ is the Aharonov-Bohm
phase due to the magnetic flux threading through the AB ring.

 Thermal current can be derived in the same way. The only difference between 
electric and thermal currents is the multiplying factor in the integrands
of the Landauer-B\"{u}ttiker formula. The charge of an electron, $-e$, in the 
electric current is replaced by the energy of an electron, $\epsilon - \mu_L$,
in the thermal current leaving the left lead. The thermal current 
is given by the expression,
\begin{eqnarray}
Q_L &=& \frac{2}{h} \int d\epsilon ~ (\epsilon - \mu_L) ~T(\epsilon)~
      [f_L(\epsilon) - f_R (\epsilon)].
\end{eqnarray}
Here $T(\epsilon)$ is the transmission probability which we derived in 
the above. 
 
\end{widetext}

\section{Wide conduction band limit}\label{wideband}
 In a wide conduction band limit, all the Green's functions related to the 
leads become energy-independent constants. 
Especially, $g_p^{r,a} (\epsilon) = \mp i$, $\overline{N}_p(\epsilon) = 1$,
and $D^{r,a} (\epsilon) = 1 + \gamma$. 
The expression of the current can be highly simplified in this case.
All the relevant Green's functions of two leads are given by the equations,
\begin{subequations}
\begin{eqnarray}
G_p^{r,a} (\epsilon) &=& \mp i\pi N_p, \\
G_{p}^{<} (\epsilon) &=& 2\pi N_p f_p(\epsilon), \\
\overline{G}_p^{r,a} (\epsilon) 
 &=& \mp \frac{i\pi N_p }{ 1 + \gamma}, \\
\overline{G}_{p}^{<} (\epsilon) 
 &=& \frac{2\pi N_p }{ (1+\gamma)^2} 
    \left[ f_p(\epsilon) + \gamma f_{\bar{p}}(\epsilon) \right], \\
G_{p\bar{p}0}^{r,a} 
 &=& - \frac{\pi^2 N_L N_R }{ 1 + \gamma} ~T_{p\bar{p}}, \\
G_{p\bar{p}0}^{<} 
 &=& \frac{2i\pi^2 N_L N_R }{ (1+\gamma)^2} ~T_{p\bar{p}} 
   \left[ f_p(\epsilon) - f_{\bar{p}} (\epsilon) \right].
\end{eqnarray} 
\end{subequations}
The parameters introduced in the previous Appendix are also simplified,
\begin{subequations}
\begin{eqnarray}
Z_L^{r} 
 &=& \frac{1}{ (1 + \gamma)^2} \left[ \Gamma_L - \gamma \Gamma_R - i(z+z^*) \right], \\
Z_R^{r} 
 &=& \frac{1}{ (1 + \gamma)^2} \left[ \Gamma_R - \gamma \Gamma_L - i(z+z^*) \right], \\
\overline{\Gamma}_L 
 &=& \frac{1}{ (1 + \gamma)^2} \left[ \Gamma_L + \gamma \Gamma_R - i(z-z^*) \right], \\
\overline{\Gamma}_R 
 &=& \frac{1}{ (1 + \gamma)^2} \left[ \Gamma_R + \gamma \Gamma_L + i(z-z^*) \right].
\end{eqnarray} 
\end{subequations}
Finally, the parameters in the transmission spectral function become
\begin{widetext}
\begin{eqnarray}
Z_{LR} &=& \frac{4}{ \Gamma (1+ \gamma)^3} 
   \left[ \Gamma_L - \gamma \Gamma_R - i(z+z^*) \right]
   \left[ \Gamma_R - \gamma \Gamma_L - i(z+z^*) \right]  \nonumber\\
 &=& - \overline{\Gamma} \left[ T_0 - g(1 - T_0\cos^2\phi) 
   + 2i \sqrt{gT_0(1-T_0)} \cos\phi  \right].
\end{eqnarray} 
Here $\overline{\Gamma} = \Gamma/(1+\gamma)$ and $\Gamma = \Gamma_L + \Gamma_R$.
After some algebra, we find $\mbox{Im} G_{IL}^{<}$ to be 
$\mbox{Im} G_{IL}^{<} (\epsilon)
 = T (\epsilon) ~\left[ f_R(\epsilon) - f_L (\epsilon) \right]$ with 
\begin{eqnarray}
T (\epsilon) 
 &=& \frac{4\gamma }{ (1+\gamma)^2} 
   + \frac{4 \overline{\Gamma}_L \overline{\Gamma}_R  }
      { \overline{\Gamma}_L +\overline{\Gamma}_R } ~\mbox{Im} G_d^{r}
   - 4 \frac{1-\gamma }{ 1 + \gamma} ~
     \frac{1 }{ \overline{\Gamma}_L +\overline{\Gamma}_R }~
     \mbox{Im} \left\{ G_d^{r} \left[ \overline{\Gamma}_L Z_R^* 
        + \overline{\Gamma}_R Z_L^* \right] \right\}.
\end{eqnarray}
\end{widetext}
The transmission spectral function $T(\epsilon)$ can be written in another form,
\begin{eqnarray}
T (\epsilon) 
 &=& T_0 + 2 \overline{\Gamma} \sqrt{g T_0 (1-T_0)} ~\cos\phi ~\mbox{Re} G_d^r 
    \nonumber\\
 &&  + \overline{\Gamma} \left[ T_0 - g(1 - T_0\cos^2\phi) \right]~
     \mbox{Im} G_d^r, \\
T_0 &=& \frac{4\gamma }{ (1+\gamma)^2}, ~~~
  g ~=~ \frac{4\Gamma_L \Gamma_R }{ ( \Gamma_L + \Gamma_R)^2 }.     
\end{eqnarray}
$T_0$ and $g$ are the dimensionless conductance of direct tunneling
and a resonant tunneling through a quantum dot, respectively. 
We note that the expression of $T(\epsilon)$ in the wide conduction band limit
was derived in the literature.\cite{kimselmanPRL, bulka, hofstetter}

\end{document}